\documentclass[11pt]{article}
\usepackage{geometry}

\usepackage{graphics,graphicx,color,float, hyperref}
\usepackage{epsfig,bbm}
\usepackage{bm}
\usepackage{mathtools}
\usepackage{amsmath}
\usepackage{amssymb}
\usepackage{amsfonts}
\usepackage{amsthm}
\usepackage {times}
\usepackage{layout}
\usepackage{setspace}
\usepackage{tikz-cd}
\allowdisplaybreaks[3]

\newtheorem{fact}{Fact}

\newcommand{\beq}{\begin{equation}}
\newcommand{\enq}{\end{equation}}
\newcommand{\bel}{\begin{lemma}}
\newcommand{\enl}{\end{lemma}}
\newcommand{\bet}{\begin{theorem}}
\newcommand{\ent}{\end{theorem}}

\newcommand{\err}{\mathrm{err}}
\newcommand{\eps}{\varepsilon}
\newcommand{\seps}{\sqrt{\varepsilon}}
\newcommand{\dinfty}{\ensuremath{\mathrm{D}_{\infty}}}
\newcommand{\dinftyeps}{\ensuremath{\mathrm{D}_{\infty}^{\eps}}}
\newcommand{\dseps}[1]{\ensuremath{\mathrm{D}_s^{#1}}}
\newcommand{\dzeroseps}[1]{\ensuremath{\mathrm{D}_{\mathrm{H}}^{#1}}}

\newcommand*{\cJ}{\mathrm{J}}
\newcommand*{\cC}{\mathrm{C}}
\newcommand*{\cA}{\mathcal{A}}

\newcommand*{\cM}{\mathcal{M}}

\newcommand*{\cB}{\mathcal{B}}

\newcommand*{\cK}{\mathcal{K}}
\newcommand*{\cN}{\mathcal{N}}

\newcommand*{\cX}{\mathcal{X}}

\newcommand*{\cZ}{\mathcal{Z}}
\newcommand*{\cE}{\mathcal{E}}

\newcommand{\cP}{\mathcal{P}}
\newcommand{\Opt}{\mathrm{Opt}}
\newcommand{\Br}{\mathrm{BR}}
\newcommand{\Ext}{\mathrm{Ext}}
\newcommand{\bOpt}{\mathrm{\textbf{Opt}}}
\newcommand{\bBr}{\mathrm{\textbf{BR}}}
\newcommand{\bExt}{\mathrm{\textbf{Ext}}}

\newcommand{\supp}{\mathrm{supp}}
\newcommand{\suppress}[1]{}
\newcommand{\drawn}{\leftarrow}
\newcommand{\defeq}{\ensuremath{ \stackrel{\mathrm{def}}{=} }}

\newcommand {\br} [1] {\ensuremath{ \left( #1 \right) }}

\newcommand {\minusspace} {\: \! \!}
\newcommand {\smallspace} {\: \!}
\newcommand {\fn} [2] {\ensuremath{ #1 \minusspace \br{ #2 } }}

\newcommand {\relent} [2] {\fn{\mathrm{D}}{#1 \middle\| #2}}
\newcommand {\dmax} [2] {\fn{\mathrm{D}_{\max}}{#1 \middle\| #2}}
\newcommand {\mutinf} [2] {\fn{\mathrm{I}}{#1 \smallspace : \smallspace #2}}
\newcommand {\trimutinf} [3] {\fn{\mathrm{I}}{#1 \smallspace : \smallspace #2 : \smallspace #3}}

\newcommand {\condmutinf} [3] {\mutinf{#1}{#2 \smallspace \middle\vert \smallspace #3}}

\newcommand {\dheps} [3] {\ensuremath{\mathrm{D}_{\mathrm{H}}^{#3}\left(#1 \| #2\right)}}
\newcommand {\id} {\ensuremath{\mathrm{I}}}
\newcommand {\dstarft} {\ell}

\newcommand*{\cV}{\mathcal{V}}
\newcommand*{\cY}{\mathcal{Y}}

\mathchardef\mhyphen="2D

\newcommand*{\good}{\mathrm{Good}}

\makeatletter
\newcommand*{\rom}[1]{\expandafter\@slowromancap\romannumeral #1@}
\makeatother

\mathchardef\mhyphen="2D

\newtheorem{definition}{Definition}

\newtheorem{theorem}{Theorem}
\newtheorem{lemma}{Lemma}
\newtheorem{corollary}{Corollary}

\newenvironment{proofof}[1]{\noindent{\bf Proof of #1:}}{\qed\\}

\begin {document}
\title{A unified approach to source and message compression}
\author{
Anurag Anshu\footnote{Institute for Quantum Computing and Department of Combinatorics and Optimization, University of Waterloo, and Perimeter Institute for Theoretical Physics, \texttt{aanshu@uwaterloo.ca}} \qquad
Rahul Jain\footnote{Center for Quantum Technologies, National University of Singapore and MajuLab, UMI 3654, 
Singapore. \texttt{rahul@comp.nus.edu.sg}} \qquad 
Naqueeb Ahmad Warsi\footnote{Center for Quantum Technologies, National University of Singapore and School of Physical and Mathematical Sciences, Nanyang Technological University, Singapore and IIITD, Delhi. \texttt{warsi.naqueeb@gmail.com}} 
}
\date{}
\maketitle

\abstract{ We study the problem of source and message compression in the one-shot setting for the point-to-point and multi-party scenarios (with and without side information).  We derive achievability results for these tasks in a unified manner, using the techniques of {\em convex-split}, which was introduced in \cite{AnshuDJ14} and {\em position-based decoding} introduced in~\cite{AnshuJW17}, which in turn uses {\em hypothesis testing} between distributions. These results are in terms of smooth max divergence and smooth hypothesis testing divergence.  As a by-product of the tasks studied in this work, we obtain several known source compression results (originally studied in the asymptotic and i.i.d. setting) in the one-shot case. 

One of our achievability results includes the problem of message compression with side information, originally studied in \cite{BravermanRao11}. We show that both our result and the result in \cite{BravermanRao11} are near optimal in the one-shot setting by proving a converse bound. 

}

\section{Introduction}
\label{intro}

\suppress{
Source compression is a fundamental task in information theory first studied by Shannon in his landmark paper~\cite{Shannon}.  This task was later extended to various network settings for example by Slepian and Wolf \cite{SlepianW73}, Wyner \cite{Wyner75}  and Wyner and Ziv \cite{WynerZ76}. These works considered the asymptotic, independent and identically distributed (i.i.d.)  setting. 

In this work we consider source and message compression in various network communication scenarios and present a unified approach to arrive at communication bounds. Message compression is a task when a random variable correlated with the source is sought to be sent with low communication. We start with a one-sender-one-receiver task. We then consider a two-senders-one-receiver task followed by a one-sender-two-receivers task. We combine these two to consider a two-senders-two-receivers task. 

We present our communication bounds in the one-shot setting which imply optimal bounds for these tasks in the asymptotic i.i.d setting. One-shot information theory has been studied extensively in the recent years both in the classical and quantum models. Apart from being practically  relevant (since there is no i.i.d. assumption) it often provides interesting new insights and conceptual advances into the working and design of communication protocols, as the complications and conveniences of the i.i.d assumption are not present. One-shot information theory has been particularly useful in communication complexity while dealing with the important and consequential {\em direct sum, direct product} and {\em composition} questions. Answering these questions has applications  in computational complexity as well. 

As applications of our results we reproduce several known results in network communication theory both in the one-shot and i.i.d. settings, further exhibiting  the power of our unified framework. 

There are two main techniques that we use to arrive at our results. First is the {\em convex-split } technique, which was introduced in \cite{AnshuDJ14} for a related problem in the quantum domain. Convex-split technique is closely related to the well known {\em rejection sampling } technique, used in various information theoretic tasks in several works~\cite{Jain:2003, Jain:2005, HJMR10, BravermanRao11}. The other technique that we use is {\em position-based decoding} introduced in~\cite{AnshuJW17}, which in turn uses {\em hypothesis testing} between distributions. These two techniques used together allow us to construct all our protocols. 

\subsection*{Our results} We start with the following one-sender-one-receiver task. For all our results in this section let $\eps > 0$ be a sufficiently small constant which represents an error parameter\footnote{We do not attempt to optimize constants appearing in this paper.}.  

\vspace{0.1in} 

\noindent {\bf Task 1: One-sender-one-receiver message compression with side information at the receiver.}
\label{task1}
There are two parties Alice and Bob. Alice possesses random variable $X$, taking values over a finite set $\cX$ (all sets that we consider in this paper are finite) and  a random variable $M$, taking values over a set $\cM$. Bob possesses random variable $Y$, taking values over a set $\cY$ such that $M$ and $Y$ are independent given $X$ represented by $M-X-Y$. Alice sends a message to Bob and at the end Bob outputs random variable $\hat{M}$ such that  $\|p_{XYM} - p_{XY\hat{M}}\| \leq O(\seps)$, where $\|.\|$ is the $\ell_1$ norm. They are allowed to use shared randomness between them which is independent of $XYM$ at the beginning of the protocol.  

\vspace{0.1in} 

This task is particularly relevant from the point of view of communication complexity, where $(X,Y)$ can be viewed as inputs given to Alice and Bob respectively from a prior distribution and $M$ can be viewed as the message Alice wants to send to Bob. It was studied in~\cite{Jain:2003, HJMR10} when the distribution of $(X,Y)$ is product and in~\cite{BravermanRao11} for general $(X,Y)$. Here, we present a new protocol for this task using the aforementioned techniques of convex split and position-based decoding and show the following achievability result. 
\suppress{In Section \ref{sec:classicalcompression}, we provide a close comparison between our protocol and the protocol presented by~\cite{BravermanRao11}. We show the optimality of both these protocols and discuss close relationship between them, notwithstanding the fact that they are not related to each other in an obvious way. It may be noted that Braverman and Rao  considered the expected communication cost required for above task (same holds for the work \cite{HJMR10}), whereas we consider worst case communication cost. Our discussion in Section \ref{sec:classicalcompression} also develops around the worst case communication cost of the protocol from \cite{BravermanRao11}, which we state as Fact \ref{lembrraoprot}.}

\begin{theorem}[Achievability for Task 1]
\label{extensionachievability}
Let $\delta \geq 0$. Let $R$ be a natural number such that,
\begin{equation*} 
R \geq \min_{\substack{(\tilde{X}, \tilde{Y}, \tilde{M}, T,E): \\ \|p_{\tilde{X}\tilde{Y}\tilde{M}} -p_{XYM}\|\leq \delta  \\ \tilde{Y}-\tilde{X}-\tilde{M}E}} \left(\dseps{\eps}(p_{\tilde{X}\tilde{M}E} \| p_{\tilde{X}} \times p_T) - \dzeroseps{\eps}(p_{\tilde{Y}\tilde{M}E} \| p_{\tilde{Y}} \times p_T) + O\left(\log \frac{1}{\eps}\right) \right) ,
\end{equation*}
where  $E$ takes values over a set $\cE$ and $T$ takes values over set $\cE \times \cM$. There exists a shared randomness assisted protocol in which Alice communicates $R$ bits to Bob and Bob outputs random variable $\hat{M}$ satisfying $\|p_{XYM} - p_{XY\hat{M}}\|\leq \delta + O(\seps)$. 
\end{theorem}
Please refer to Section~\ref{sec:prelim} for the definitions of $\dseps{\eps}(\cdot)$ and $\dzeroseps{\eps}(\cdot)$. The minimization above over $\tilde{X}, \tilde{Y}, \tilde{M}$ and $E$ (which we refer to as {\em extension} of $\tilde{M}$) and  $T$ (which is used in shared randomness) may potentially decrease the amount of communication between Alice and Bob. In our converse result below, we show that this is indeed the case. 
\begin{theorem}[Converse for Task 1] \label{extensionconverse}
Any communication protocol for Task $1$ must satisfy:
$$ R \geq \min_{\substack{(\tilde{X}, \tilde{Y}, \tilde{M}, U, E): \\ \|p_{\tilde{X}\tilde{Y}\tilde{M}} -p_{XYM}\|\leq 3\seps  \\ \tilde{Y}-\tilde{X}-\tilde{M}E}} \left(\dseps{3\seps}(p_{\tilde{X}\tilde{M}E} \| p_{\tilde{X}} \times p_U) - \dzeroseps{3\seps}(p_{\tilde{Y}\tilde{M}E} \| p_{\tilde{Y}} \times p_U) - O\left(\log \frac{1}{\eps}\right) \right) ,$$
where $R$ is the communication (in bits) between Alice and Bob, $E$ (taking values in $\cE$) is a specific extension (defined subsequently in the proof of this result) of $\tilde{M}$ and $U$ is uniformly distributed over $\cM \times \cE$.  
\end{theorem}

}

Source compression is a fundamental task in information theory first studied by Shannon in his landmark paper~\cite{Shannon}.  This task was later extended to various network settings for example by Slepian and Wolf \cite{SlepianW73}, Wyner \cite{Wyner75}  and Wyner and Ziv \cite{WynerZ76}. These works considered the asymptotic, independent and identically distributed (i.i.d.)  setting. 

Compression protocols have been particularly relevant in communication complexity \cite{Yao79, Kushilevitz96}, where Alice and Bob wish to compute a joint function of their inputs $x,y$ (that are sampled from a joint distribution $p_{XY}$). Upon receiving her input, Alice sends a message $M$ to Bob, who sends the next message to Alice conditioned on Alice's message and his input. This process continues till both parties have computed the desired function up to some error. Observe that $M$ and $Y$ are independent conditioned on $X$. An important task in communication complexity is to communicate $M$ with small communication (referred to as~\textit{message compression}, see Task \ref{task1} below), which has been investigated by several works~\cite{Jain:2003, Jain:2005, HJMR10, BravermanRao11}. This is connected to important and fundamental questions in communication complexity, namely~\textit{direct sum, direct product and composition}, which relate the resource requirements for many independent instances of a task to the resource requirement of a single instance of the same task.  

In this work we consider source and message compression in various network communication scenarios and present a unified approach to arrive at communication bounds. Starting from a one-sender-one-receiver task, we consider a two-senders-one-receiver task followed by a one-sender-two-receivers task. These tasks are summarized in Figure \ref{fig:alltasks}. We combine these two to consider a two-senders-two-receivers task. It can be observed that our approach extends to more complicated network scenarios as well. We particularly focus on message compression in network scenarios due to growing interest in the problems related to multi-party communication complexity \cite{ChandraFL83, Sherstov12, LeeS09, BeameH12}.

We present our communication bounds in the one-shot setting and sketch how these bounds behave in the asymptotic i.i.d setting. We leave the question of second order and asymptotic non-i.i.d. analysis of many of these results to future work (second order and asymptotic non-i.i.d. analysis of some of the results has already been achieved in known literature). One-shot information theory has been studied extensively in the recent years both in the classical and quantum models. Apart from being practically  relevant (since there is no i.i.d. assumption) it often provides interesting new insights and conceptual advances into the working and design of communication protocols, as the complications and conveniences of the i.i.d assumption are not present. One-shot information theory has been particularly useful in communication complexity while dealing with the aforementioned  direct sum, direct product and composition questions. 

As applications of our results we reproduce several known results in network communication theory both in the one-shot and i.i.d. settings, further exhibiting the power of our unified framework. 

There are two main techniques that we use to arrive at our results. First is the {\em convex-split } technique, which was introduced in \cite{AnshuDJ14} for a related problem in the quantum domain. Convex-split technique is closely related to the well known {\em rejection sampling } technique, used in various information theoretic tasks in several works~\cite{Jain:2003, Jain:2005, HJMR10, BravermanRao11, RadhakrishnanSW16}. The other technique that we use is {\em position-based decoding} introduced in~\cite{AnshuJW17}, which in turn uses {\em hypothesis testing} between distributions. These two techniques used together allow us to construct all our protocols. For precise definition of all the information theoretic quantities appearing in this section, please refer to Section \ref{sec:prelim}.

\begin{itemize}
\item \textit{Convex-split technique:} Central to this technique is the \textit{convex-split lemma} \cite{AnshuDJ14}, which is a statement of the following form. Let $p_{AB}$ be a probability distribution over the set $\cA\times \cB$, $p_{B'}$ be a probability distribution (possibly different from $p_B$) over the set $\cB$. Define a probability distribution $p_{AB_1B_2\ldots B_{2^R}}$ as
$$p_{AB_1B_2\ldots B_{2^R}}(a,b_1,b_2,\ldots b_{2^R}) := \frac{1}{2^R}\sum_j p_{AB}(a,b_j)\cdot p_{B'}(b_1)\cdot \ldots p_{B'}(b_{j-1})\cdot p_{B'}(b_{j+1})\ldots \cdot p_{B'}(b_{2^R}).$$
Then 
$$\relent{p_{AB_1B_2\ldots B_{2^R}}}{p_A\times p_{B'}\times p_{B'}\times \ldots p_{B'}} \leq \eps,$$ if $R \geq \dmax{p_{AB}}{p_A\times p_{B'}} + \log\frac{1}{\eps}$. Here $\relent{p_X}{p_{X'}} = \sum_x p_X(x)\log \frac{p_X(x)}{p_{X'}(x)}$ is the relative entropy and $\dmax{p_X}{p_{X'}} = \max_x\log \frac{p_X(x)}{p_{X'}(x)}$ is the max divergence. 

In this work, we shall use a corollary of above result, which is a statement of the form 
\begin{equation} \label{eq:convex}
\frac{1}{2}\| p_{AB_1B_2\ldots B_{2^R}} - p_A\times p_{B'}\times p_{B'}\times \ldots p_{B'}\|_1 \leq \eps + \delta,
\end{equation} if $R\geq \dseps{\eps}(p_{AB}\| p_A\times p_{B'}) + 2\log\frac{3}{\delta}$ and $\dseps{\eps}(p_X\| p_{X'})$ is the information spectrum relative entropy. Convex-split lemma is reminiscent of \cite[Lemma 4.1]{JainSWZ13}, which was also independently obtained as the \textit{soft-covering lemma} in \cite{Cuff13} (see also \cite{WatanabeKT15, GoldfeldCP16, SongCP15} for applications), but there are two points of difference. First is that the former is in terms of relative entropy, whereas the latter is in terms of the variational distance. Second, convex-split lemma accommodates the random variable $B'$ which is not related to the random variable $B$, a feature that is not present in the soft-covering lemma. In fact, this feature is essential for our protocol, as we shall show various optimality results using the fact that we can construct protocols using an arbitrary random variable $B'$. 

\item \textit{Position-based decoding technique:} This technique uses hypothesis testing to locate the index $j$ where correlation between random variables $A$ and $B$ is according to $p_{AB}$ in the distribution $p_{AB_1B_2\ldots B_{2^R}}(a,b_1,b_2,\ldots b_{2^R})$ defined above. The technique succeeds with error $\eps+3\delta$ as long as $R \leq \dheps{p_{AB}}{p_A\times p_{B'}}{\eps} + \log \delta$. Here $\dheps{.}{.}{\eps}$ is the smooth hypothesis testing divergence.
\end{itemize} 

These two techniques are dual and  complementary to each other. One is about diluting correlation and the other is about recovering correlation. 

\begin{figure}[ht]
\centering
\begin{tikzpicture}[xscale=0.6,yscale=0.7]

\begin{scope}
%Onesenderonereceiver

\draw[ultra thick] (-8,5.5) rectangle (0,2.5);
\draw[ultra thick] (8,5.5) rectangle (0,2.5);

\node at (-7, 5) {Alice};
\node at (-1,5) {Bob};

\node at (-5.5, 4.3) {$X$};
\node at (-2.5, 4.3) {$Y$};
\node at (-6.5, 3.3) {$M$};

\draw (-5.7, 4.1) -- (-6.2, 3.6);
\draw (-5.3, 4.3) -- (-2.8, 4.3);

\node at (1, 5) {Alice};
\node at (7,5) {Bob};

\node at (2.5, 4.3) {$X$};
\node at (5.5, 4.3) {$Y$};
\node at (5, 3.3) {$M$};

\draw (2.7, 4.3) -- (5.2, 4.3);
\draw (2.7, 4.1) -- (4.7, 3.5);

\node at (0, 2.0) {\small Task $1$: One sender, one receiver with side information.};
\node at (0,1.4) {\small Its special case is Task $7$: lossy source compression.};

\end{scope}

\begin{scope}[yshift=-4.8cm]
%Twosenderonereceiver

\draw[ultra thick] (-8,5.5) rectangle (0,0);
\draw[ultra thick] (8,5.5) rectangle (0,0);

\node at (-7, 5) {Alice};
\node at (-1.1,5) {Charlie};
\node at (-7, 0.5) {Bob};

\node at (-5.5, 3.3) {$X$};
\node at (-2.5, 2.7) {$Z$};
\node at (-6.5, 4.3) {$M$};
\node at (-5.5, 2.2) {$Y$};
\node at (-6.5, 1.2) {$N$};

\draw (-5.7, 3.5) -- (-6.2, 4.1);
\draw (-5.3, 3.3) -- (-2.8, 2.75);
\draw (-5.5,3.1) -- (-5.5, 2.5);
\draw (-5.3, 2.2) -- (-2.8, 2.65);
\draw (-5.7, 2.0) -- (-6.2, 1.4);

\node at (1, 5) {Alice};
\node at (6.9,5) {Charlie};
\node at (1, 0.5) {Bob};

\node at (2.5, 3.3) {$X$};
\node at (5.5, 2.7) {$Z$};
\node at (5, 4.3) {$M$};
\node at (2.5, 2.2) {$Y$};
\node at (5, 1.2) {$N$};

\draw (2.8, 3.5) -- (4.7, 4.1);
\draw (2.7, 3.3) -- (5.2, 2.75);
\draw (2.5,3.1) -- (2.5, 2.5);
\draw (2.7, 2.2) -- (5.2, 2.65);
\draw (2.7, 2.0) -- (4.7, 1.3);

\node at (0,-0.5) {\small Task $3$: Two senders, one receiver with side information.};
\node at (0,-1.1) {\small Its special case is Task $2$: Two senders, one receiver (without side information).};
\node at (0,-1.7) {\small Special cases of Task $2$ are Task $8$ (first studied by Slepian and Wolf \cite{SlepianW73})};
\node at (0,-2.3) {\small and Task $9$ (first studied by Wyner \cite{Wyner75}).};
\end{scope}

\begin{scope}[yshift=-13.2cm]
%Twosenderonereceiver

\draw[ultra thick] (-8,5.5) rectangle (0,0);
\draw[ultra thick] (8,5.5) rectangle (0,0);

\node at (-7, 5) {Alice};
\node at (-1.1,5) {Charlie};
\node at (-1, 0.5) {Bob};

\node at (-2.5, 3.3) {$Y$};
\node at (-5.5, 2.7) {$X$};
\node at (-2.5, 2.2) {$Z$};
\node at (-6.5, 3.7) {$M$};
\node at (-6.5, 1.8) {$N$};

\draw (-5.7, 2.9) -- (-6.2, 3.6);
\draw (-5.3, 2.75) -- (-2.8, 3.3);
\draw (-2.5,3.1) -- (-2.5, 2.5);
\draw (-5.3, 2.65) -- (-2.8, 2.2);
\draw (-5.7, 2.5) -- (-6.2, 2.0);
\draw (-6.5, 3.4) -- (-6.5, 2.1);

\node at (1, 5) {Alice};
\node at (6.9,5) {Charlie};
\node at (7, 0.5) {Bob};

\node at (5.5, 3.3) {$Y$};
\node at (2.5, 2.7) {$X$};
\node at (5.5, 2.2) {$Z$};
\node at (6.5, 4.2) {$M$};
\node at (6.5, 1.3) {$N$};

\draw (2.8, 2.9) -- (6.2, 4.2);
\draw (2.7, 2.75) -- (5.2, 3.3);
\draw (5.5,3.1) -- (5.5, 2.5);
\draw (2.7, 2.65) -- (5.2, 2.2);
\draw (2.8, 2.5) -- (6.2, 1.3);
\draw (6.5,3.9) -- (6.5, 1.6);

\node at (0,-0.5) {\small Task $5$: One sender, two receivers with side information.};
\node at (0,-1.1) {\small Its special case is Task $4$: One sender, two receivers (without side information).};
\end{scope}

\end{tikzpicture}
\caption{Various communication tasks considered in this work. The lines describe the conditional independence (random variables involved in a triangle are arbitrarily correlated, see the formal definition in Introduction). Task $6$ is a further generalization, not depicted above.}
 \label{fig:alltasks}
\end{figure}

\subsection*{Our results} We start with the following one-sender-one-receiver task. For all our results in this section let $\eps, \delta > 0$ be sufficiently small constants which represent error parameters.

\vspace{0.1in} 

\noindent {\bf Task 1: One-sender-one-receiver message compression with side information at the receiver.}
\label{task1}
There are two parties Alice and Bob. Alice possesses random variable $X$, taking values over a finite set $\cX$ (all sets that we consider in this paper are finite). Bob possesses random variable $Y$, taking values over a set $\cY$. Let $M$ be a random variable, taking values over a set $\cM$ and jointly distributed with $(X,Y)$, such that $M$ and $Y$ are independent given $X$ represented by $M-X-Y$. Alice sends a message to Bob and at the end Bob outputs random variable $\hat{M}$ such that  $\frac{1}{2}\|p_{XYM}- p_{XY\hat{M}}\| \leq \eps$, where $\|.\|$ represents the $\ell_1$ norm. They are allowed to use shared randomness between them which is independent of $XYM$ at the beginning of the protocol.  

\vspace{0.1in} 

As discussed earlier, this task is particularly relevant from the point of view of communication complexity, where $(X,Y)$ can be viewed as inputs given to Alice and Bob respectively from a prior distribution. It was studied in~\cite{Jain:2003, HJMR10} when the distribution of $(X,Y)$ is product and in~\cite{BravermanRao11} for general $(X,Y)$. All these results are in the one-shot setting. It was also studied when the random variable $Y$ is not present as a side information in the work \cite{Cuff08}, in the asymptotic and i.i.d. setting. The `worst-case' analogue of this task without side information $Y$, where the error has to be small for each $x$, is the task of \textit{channel simulation}. It was first studied in the asymptotic and i.i.d. setting in the works \cite{BennettSST02, BennettDHSW14}, under the name of the Classical Reverse Shannon Theorem. The work \cite{HJMR10} also considered its one-shot variant where the communication cost was measured in terms of the expected number of bits transmitted. 

Here, we discuss two ways of analyzing Task \ref{task1}. First is the protocol of Braverman and Rao from \cite{BravermanRao11}, who analyzed the expected communication cost of their protocol. We show that their protocol is nearly optimal in the present setting (that is the worst case communication cost or the total number of bits communicated from Alice to Bob) in the following theorem. 

\begin{theorem}[Achievability and converse for Task 1 using Braverman and Rao's protocol]
\label{BRachievabilityconverse}
Let $\eps, \delta \in (0,1)$. Let $R$ be a natural number such that,
\begin{equation*} 
R \geq \dseps{\eps}(p_{XMY}\| p_Y(p_{X\mid Y} \times p_{M\mid Y})) + 2\log\frac{1}{\delta},
\end{equation*}
where $p_Y(p_{X\mid Y} \times p_{M\mid Y})$ is the probability distribution defined as $$p_Y(p_{X\mid Y} \times p_{M\mid Y})(x,m,y) = p_Y(y)p_{X\mid Y=y}(x)p_{M\mid Y=y}(m).$$There exists a shared randomness assisted protocol in which Alice communicates $R$ bits to Bob and Bob outputs random variable $\hat{M}$ satisfying $\frac{1}{2}\|p_{XYM} - p_{XY\hat{M}}\|\leq \eps + 3\delta$. 

Further, any communication protocol for Task \ref{task1} must satisfy:
$$R \geq \dseps{\eps/ (1-\delta)}(p_{XMY}\| p_Y(p_{X\mid Y} \times p_{M\mid Y})) - \log \frac{1}{\delta},$$
where $R$ is the communication (in bits) between Alice and Bob
\end{theorem}

The proof of this result follows from Theorem \ref{main:theo}. It is not immediately clear how the protocol of Braverman and Rao generalizes to the multi-party setting. We present a new protocol for Task \ref{task1}, using the aforementioned techniques of convex-split and position-based decoding, and show a new achievability result in Theorem \ref{extensionachievability}. We are able to generalize our construction, using convex-split and position-based decoding techniques, to various multi-party settings as we show subsequently.
\begin{theorem}[Achievability for Task 1]
\label{extensionachievability}
Let $\eps, \delta \in (0,1)$. Let $R$ be a natural number such that,
\begin{equation*} 
R \geq \min_{E,T: Y-X-(M,E)} \left(\dseps{\delta}(p_{XME} \| p_{X} \times p_T) - \dzeroseps{\eps}(p_{YME} \| p_{Y} \times p_T) + 3\log \frac{3}{\delta}\right)  ,
\end{equation*}
where  $E$ takes values over a set $\cE$ and $T$ takes values over set $\cE \times \cM$. There exists a shared randomness assisted protocol in which Alice communicates $R$ bits to Bob and Bob outputs random variable $\hat{M}$ satisfying $\frac{1}{2}\|p_{XYM} - p_{XY\hat{M}}\|\leq \eps + 4\delta$. 
\end{theorem}
Please note that in the proof of this result we crucially use the fact that the convex-split lemma can accommodate random variable $B'$ different from $B$. The minimization over $E$ (which we refer to as {\em extension} of $M$) and  $T$ (which is used in shared randomness) may potentially decrease the amount of communication between Alice and Bob. In our converse result below, we show that this is indeed the case. This also establishes the near optimality of our protocol.   
\begin{theorem}[Converse for Task 1] \label{extensionconverse}
Fix a $\delta\in (0,1)$. Any communication protocol for Task \ref{task1} must satisfy:
$$ R \geq \min_{E: Y-X-(M,E)}\left(\dseps{0}(p_{XME} \| p_{X} \times p_U) - \dzeroseps{\eps/(1-\delta)}(p_{YME} \| p_{Y} \times p_U) - \log \frac{1}{\delta}\right) ,$$
where $R$ is the communication (in bits) between Alice and Bob, $E$ (taking values in $\cE$) is a specific extension (defined subsequently in the proof of this result) of $M$ and $U$ is uniformly distributed over $\cM \times \cE$.  
\end{theorem}

We highlight some important aspects of above results. 
\begin{itemize}
\item The condition $M-X-Y$ is very crucially exploited in our protocol (and in the protocols given in aforementioned works for this task). In the case where $M,X,Y$ do not form a Markov chain, one needs to optimize over a new random variable $V$ that satisfies the conditions $V-MX-Y, X-YV-M$. Owing to the lack of a better understanding of the random variable $V$, we do not pursue this case further in present work.

\item The achievability and converse results given above converge (in terms of \textit{rate} of communication) to the conditional mutual information $\condmutinf{X}{M}{Y}$ in the asymptotic and i.i.d. setting. For this, we use the asymptotic i.i.d. analysis of information spectrum relative entropy given in \cite{TomHay13} to conclude that the rate of communication is equal to $\relent{p_{XMY}}{p_Y(p_{X\mid Y}\times p_{M\mid Y})}$ which evaluates to $\condmutinf{X}{M}{Y}$ by direct analysis. 

\item The quantum analogue of Task \ref{task1} is the problem of quantum state redistribution \cite{Devatakyard, YardD09}. To motivate this analogy, observe that the Task \ref{task1} captures one round of communication in a classical communication protocol and quantum state redistribution captures one round of communication in a quantum communication protocol \cite{Dave14}. Theorems \ref{BRachievabilityconverse}, \ref{extensionachievability} and \ref{extensionconverse} give a near optimal one-shot result for Task \ref{task1}. On the other hand, a similar result for quantum state redistribution is unknown despite several recent efforts \cite{Berta14, DattaHO14, AnshuDJ14, AnshuJW17-b} and is one of the major open problems in quantum information theory.   

\item The shared randomness used in the achievability protocol in Theorem \ref{extensionachievability} is many independent copies of the random variable $T$. This requires a large number of bits of shared randomness. But it can be significantly reduced by using pairwise independent random variables, by using an equivalent version of the convex-split lemma in the classical and classical-quantum case, as discussed in \cite{AnshuJW17MC}. The same holds for all the results appearing below.  
\end{itemize}

Next we consider several generalizations of above task. We start with a two-sender-one-receiver task and then extend it to side information with the receiver. Following this, we consider a one-sender-two-receiver task and then extend it to side information with the receiver. We have divided our discussion into these tasks since each task will require the application of our techniques in a unique way. After we have discussed these tasks, it will be seen that the use of our techniques extends in a similar fashion to more complex network scenarios. 

\vspace{0.1in}

\noindent {\bf Task 2: Two-senders-one-receiver message compression.}  There are three parties Alice, Bob and Charlie. Alice holds a random variable $X$ and Bob holds a random variable $Y$. Let $(M,N)$ be a random variable pair jointly correlated with $(X,Y)$ such that $M-X-Y-N$. Alice and Bob send a message each to Charlie and at the end Charlie outputs $(\hat{M},\hat{N})$ such that $\frac{1}{2}\|p_{XYMN}-p_{XY\hat{M}\hat{N}}\| \leq \eps$. Shared randomness is allowed between Alice and Charlie and  between Bob and Charlie. 

\vspace{0.1in} 

We show the following achievability result for this task. 
\begin{theorem}[Achievability for Task 2]
\label{slepianwolfachieve}
Fix $\eps, \eta_1, \eta_2, \delta \in (0,1)$. Let $S,T$ be random variables taking values over the same sets as $M,N$ respectively. Let $R_A,R_B$ be natural numbers such that there exist natural numbers $r_A, r_B \geq 0$ which satisfy the constraints.
\begin{align*}
R_A+r_A &\geq \dseps{\eta_1}(p_{XM} \| p_X \times p_S) +    2\log \frac{3}{\delta} ,\\
R_B +r_B& \geq \dseps{\eta_2}(p_{YN} \| p_Y \times p_T) +    2\log \frac{3}{\delta},
\end{align*}
and
$$\Pr_{(m,n)\leftarrow p_{MN}}\left\{\frac{p_{MN}(m,n)}{p_S(m)p_N(n)} \leq \frac{2^{r_A}}{\delta} \quad\text{or}\quad \frac{p_{MN}(m,n)}{p_M(m)p_T(n)} \leq \frac{2^{r_B}}{\delta} \quad\text{or}\quad\frac{p_{MN}(m,n)}{p_S(m)p_T(n)} \leq \frac{2^{r_A+r_B}}{\delta}\right\} \leq \eps.$$
In particular, the following choice of $R_A, R_B$ is sufficient, where $\eps_1+\eps_2+\eps_3 \leq \eps- 3\delta$. 
\begin{align}
\label{choicetask2}
R_A &\geq \dseps{\eta_1}(p_{XM} \| p_X \times p_S) -  \dzeroseps{\eps_1}(p_{MN}\| p_{S}\times p_{N}) + 4\log \frac{3}{\delta},\nonumber\\
R_B &\geq \dseps{\eta_2}(p_{YN} \| p_Y \times p_T) - \dzeroseps{\eps_2}(p_{MN}\| p_M\times p_T)  + 4\log \frac{3}{\delta},\nonumber\\
R_A+R_B &\geq \dseps{\eta_1}(p_{XM} \| p_X \times p_S) + \dseps{\eta_2}(p_{YN} \| p_Y \times p_T) - \dzeroseps{\eps_3}(p_{MN}\| p_S\times p_T) + 6\log \frac{3}{\delta}.
\end{align}
There exists a shared randomness assisted protocol with communication $R_A$ bits from Alice to Charlie and $R_B$  bits  from Bob to Charlie, in which Charlie outputs random variable pair $(\hat{M},\hat{N})$ such that $\frac{1}{2}\|p_{XYMN} - p_{XY\hat{M} \hat{N}}\|\leq \eps+ \eta_1+\eta_2 + 8\delta$. 
\end{theorem}

\vspace{0.1in}

\noindent {\bf Remark:} We can optimize over extensions $E$ as in Theorem~\ref{extensionachievability}. However we skip explicit mention of this optimization for ease of exposition and for brevity, both in the statement above and in its proof. We do the same for all the results later in this section. 

\vspace{0.1in}

Next we consider the same task but with side information with Charlie. 

\vspace{0.1in}

\noindent {\bf Task 3: Two-senders-one-receiver message compression with side information at the receiver.} There are three parties Alice, Bob and Charlie. Alice holds a random variable $X$, Bob holds a random variable $Y$ and Charlie holds a random variable $Z$. Let $(M,N)$ be a random variable pair jointly correlated with $(X,Y,Z)$ such that $M-X-(Y,Z)$ and $N-Y-(X,Z).$  Alice and Bob send a message each to Charlie and at the end Charlie outputs $(\hat{M},\hat{N})$ such that $\frac{1}{2}\|p_{XYZMN}-p_{XYZ\hat{M}\hat{N}}\| \leq \eps$. Shared randomness is  allowed between Alice and Charlie and between Bob and Charlie. 

\vspace{0.1in} 

We show the following achievability result for this task. 
\begin{theorem}[Achievability for Task 3]
\label{swsideinf}
Fix $\eps_1,\eps_2,\eps_3,\delta \in (0,1)$. Let $S,T$ be random variables taking values over the same sets as $M,N$ respectively. Let $R_A,R_B$ be natural numbers  such that,
\begin{align*}
R_A &\geq \dseps{\delta}(p_{XM} \| p_X \times p_S) -  \dzeroseps{\eps_1}(p_{MNZ}\| p_{S}\times p_{NZ}) + 4\log \frac{3}{\delta} , \\
R_B &\geq \dseps{\delta}(p_{YN} \| p_Y \times p_T) - \dzeroseps{\eps_2}(p_{MZN}\| p_{MZ}\times p_T)  + 4\log \frac{3}{\delta} ,\\
R_A+R_B &\geq \dseps{\delta}(p_{XM} \| p_X \times p_S) + \dseps{\delta}(p_{YN} \| p_Y \times p_T) - \dzeroseps{\eps_3}(p_{MNZ}\| p_S\times p_T \times p_Z) + 6\log \frac{3}{\delta}.
\end{align*}
There exists a shared randomness assisted protocol with communication $R_A$ bits from  Alice to Charlie and $R_B$ bits  from Bob to Charlie, in which Charlie outputs random variable pair $(\hat{M},\hat{N})$ such that $\frac{1}{2}\|p_{XYZMN} - p_{XYZ\hat{M} \hat{N}}\|\leq \eps_1+\eps_2+\eps_3+13\delta$. 
\end{theorem}

\noindent {\bf Remark:} The statement of above theorem is in a less general form than Theorem~\ref{slepianwolfachieve}, as we do not use Fact \ref{fact:biposition} in its full generality. This is for the ease in the presentation of the results. 

\vspace{0.1in}

Next we consider the following one-sender-two-receivers task. 

\vspace{0.1in}

\noindent {\bf Task 4: One-sender-two-receivers message compression.}  There are three parties Alice, Bob and Charlie. Alice holds a random variable $X$. Let $(M,N)$ be a random variable pair jointly correlated with $X$. She sends a message to Bob and a message to Charlie. Bob and Charlie after receiving their respective messages, output random variables $\hat{M}$ and $\hat{N}$ respectively such that $\frac{1}{2}\| p_{XMN} - p_{X\hat{M} \hat{N}} \| \leq \eps.$  Shared randomness is allowed between Alice and Charlie and between Alice and Bob. 

\vspace{0.1in} 

We show the following achievability result for this task. 
\begin{theorem}[Achievability for Task 4]
\label{networkprob}
Let $\eps, \delta \in (0,1)$ and $\eps_1, \eps_2, \eps_3 \in (0,1)$ be such that $\eps_1+\eps_2+\eps_3\leq \eps$. Let $S,T$ be random variables taking values over the same sets as $M,N$ respectively. Let $R_B, R_C$ be natural numbers  such that,
\begin{align*}
R_B & \geq \dseps{\eps_1}(p_{XM} \| p_X\times p_S) + 2\log \frac{5}{\delta} ,\\
R_C & \geq \dseps{\eps_2}(p_{XN} \| p_X\times p_T) +  2\log \frac{5}{\delta} ,\\
R_B+R_C & \geq \dseps{\eps_3}(p_{XMN} \| p_{X} \times p_S \times p_T) + 2\log \frac{5}{\delta} .
\end{align*}
There exists a shared randomness assisted protocol with communication $R_B$ bits  from Alice to Bob and $R_C$ bits from Alice to Charlie, in which Bob outputs $\hat{M}$ and Charlie outputs $\hat{N}$  such that $\frac{1}{2}\| p_{XMN} - p_{X\hat{M} \hat{N}} \|\leq \eps + \delta.$
\end{theorem}

Next we consider the same task but with side information at the receivers. 

\vspace{0.1in}

\noindent {\bf Task 5: One-sender-two-receivers message compression with side information at receivers.}  There are three parties Alice, Bob and Charlie.  Alice holds a random variable $X$, Bob holds random variable $Y$ and Charlie holds random variable $Z$. Let $(M,N)$ be a random variable pair jointly correlated with $(X,Y,Z)$ such that $(M,N)-X-(Y,Z)$.  Alice sends a message to Bob and a message to Charlie. Bob and Charlie after receiving their respective messages, output random variables $\hat{M}$ and $\hat{N}$ respectively such that $\frac{1}{2}\| p_{XYZMN} - p_{XYZ\hat{M} \hat{N}} \| \leq \eps.$  Shared randomness is allowed between Alice and Bob and between Alice and Charlie. 

\vspace{0.1in} 

We show the following achievability result for this task. 
\begin{theorem}[Achievability for Task 5]
\label{networkprobs}
Let $\eps, \delta_1, \delta_2 \in (0,1)$ and $\eps_1, \eps_2, \eps_3 \in (0,1)$ be such that $\eps_1+\eps_2+\eps_3\leq \eps$. Let $S,T$ be random variables taking values over the same sets as $M,N$ respectively. Let $R_B, R_C$ be natural numbers  such that,
\begin{align*}
R_B & \geq \dseps{\eps_1}(p_{XM} \| p_X\times p_{S}) - \dzeroseps{\delta_1}(p_{MY}\| p_{S}\times p_{Y}) + 2\log \frac{5}{\delta_2},\\
R_C & \geq \dseps{\eps_2}(p_{XN} \| p_X\times p_T) - \dzeroseps{\delta_1}(p_{NZ}\| p_{T}\times p_{Z}) + 2\log \frac{5}{\delta_2} ,\\
R_B+R_C & \geq \dseps{\eps_3}(p_{XMN} \| p_{X} \times p_S \times p_T) -  \dzeroseps{\delta_1}(p_{MY}\| p_{S}\times p_{Y})-\dzeroseps{\delta_1}(p_{NZ}\| p_{T}\times p_{Z}) + 2\log \frac{5}{\delta_2}.
\end{align*}
There exists a shared randomness assisted protocol with communication $R_B$ bits  from Alice to Bob and $R_C$ bits from Alice to Charlie, in which Bob outputs $\hat{M}$ and Charlie outputs $\hat{N}$  such that $\frac{1}{2}\| p_{XYZMN} - p_{XYZ\hat{M} \hat{N}} \|\leq \eps + 2\delta_1+ 5\delta_2.$
\end{theorem}

Finally we consider the following task with two senders and two receivers. 

\vspace{0.1in}

\noindent {\bf Task 6: Two-senders-two-receivers message compression with side information at the receivers.}  There are four parties Alice, Dave, Bob and Charlie. Alice holds random variables  $(X_1,M_{11},M_{12}),$ Dave holds random variables 
$(X_2,M_{21},M_{22})$, Bob holds random variable $Y_1$ and Charlie holds random variable $Y_2$ such that $(M_{11},M_{12})-X_1-(Y_1,Y_2,X_2)$ and $(M_{21},M_{22})-X_2-(Y_1,Y_2,X_1).$  Alice sends a message each to Bob and Charlie and Dave sends a message each to Bob and Charlie. At the end Bob outputs $(\hat{M}_{11}, \hat{M}_{21})$ and Charlie outputs $(\hat{M}_{12}, \hat{M}_{22})$ such that,
$$\frac{1}{2}\|p_{X_1M_{11}M_{12} X_2M_{21}M_{22} Y_1 Y_2}-p_{X_1\hat{M}_{11}\hat{M}_{12}X_2\hat{M}_{21}\hat{M}_{22} Y_1Y_2}\| \leq \eps.$$ 
Shared randomness is allowed between pairs (Alice, Bob), (Alice, Charlie), (Dave, Bob) and (Dave, Charlie).

\vspace{0.1in}

We obtain the following achievability result for this task using arguments similar to the arguments used in obtaining previous achievability results. We skip its proof for brevity. 
\begin{theorem}[Achievability for Task 6]
Let $R^{(1)}_1, R^{(1)}_2, R^{(2)}_{1}, R^{(2)}_{2}$ be natural numbers such that for $i,j \in \{1,2\}$, 
\begin{align*}
R^{(i)}_j & \geq \dseps{\eps}(p_{X_iM_{ij}} \| p_{X_i}\times p_{M_{ij}}) - \dzeroseps{\eps}(p_{M_{ij}Y_j}\| p_{M_{ij}}\times p_{Y_j}) + 10\log \frac{1}{\delta} ,
\end{align*}
for  $i,j,k,l \in \{1,2\}$ such that $i\neq k$ or $j\neq l$,
\begin{align*}
R^{(i)}_j +R^{(k)}_l& \geq  \dseps{\eps}(p_{X_iM_{ij}} \| p_{X_i}\times p_{M_{ij}})+ \dseps{\eps}(p_{X_kM_{kl}} \| p_{X_k}\times p_{M_{kl}})\\
&\hspace{3mm} - \dzeroseps{\eps}(p_{M_{ij}Y_j}\| p_{M_{ij}} \times p_{Y_j})-\dzeroseps{\eps}(p_{M_{kl}Y_l}\| p_{M_{kl}}\times \times p_{Y_l}) + 10\log \frac{1}{\delta},
\end{align*}
for $i,j,k,l\in \{1,2\}$ such that $i\neq k$ and $ j\neq l$,
\begin{align*}
R^{(i)}_j+R^{(i)}_l+R^{(k)}_j & \geq \dseps{\eps}(p_{X_iM_{ij}M_{il}} \| p_{X_i} \times p_{M_{ij}} \times p_{M_{il}})  + \dseps{\eps}(p_{X_kM_{kj}} \| p_{X_k}\times p_{M_{kj}})\\
&\hspace{3mm} -\dzeroseps{\eps}(p_{M_{ij}M_{kj}Y_j}\| p_{M_{ij}}\times p_{M_{kj}} \times p_{Y_j})-  \dzeroseps{\eps}(p_{M_{il}Y_l}\| p_{M_{il}}\times p_{Y_l})+ 10\log \frac{1}{\delta} ,
\end{align*}
and,
\begin{align*}
R^{(1)}_1+R^{(1)}_2+R^{(2)}_1+R^{(2)}_2&\geq \dseps{\eps}(p_{X_1M_{11}M_{12}} \| p_{X_1} \times p_{M_{11}} \times p_{M_{12}})+\dseps{\eps}(p_{X_2M_{21}M_{22}} \| p_{X_2} \times p_{M_{21}} \times p_{M_{22}}) \\
&\hspace{3mm} -\dzeroseps{\eps}(p_{M_{11}M_{21}Y_1}\| p_{M_{11}}\times p_{M_{21}} \times p_{Y_1})-\dzeroseps{\eps}(p_{M_{12}M_{22}Y_2}\| p_{M_{12}}\times p_{M_{22}} \times p_{Y_2}) + 10\log \frac{1}{\delta}.
\end{align*}
There exists a shared randomness assisted protocol with communication $R^{(1)}_1$ bits from Alice to Bob, $R^{(1)}_2$ bits from Alice to Charlie, $R^{(2)}_1$ bits from Dave to Bob and $R^{(2)}_2$ bits from Dave to Charlie such that Bob outputs $(\hat{M}_{11}, \hat{M}_{21} )$ and Charlie outputs $(\hat{M}_{12}, \hat{M}_{22} )$ satisfying  $$\frac{1}{2}\|p_{X_1M_{11}M_{12} X_2M_{21}M_{22} Y_1 Y_2}-p_{X_1\hat{M}_{11}\hat{M}_{12}X_2\hat{M}_{21}\hat{M}_{22} Y_1Y_2}\|\leq 20\eps + 30\delta.$$
\end{theorem}
We state without giving further details, that the task above can be extended in a natural fashion to obtain an analogous task for multiple senders and multiple receivers and analogous communication bounds can be obtained using similar arguments.

\subsection*{Applications of our results}
Here we consider several tasks studied in previous works and show that our results imply the results shown in these works.  Consider the following task.
\vspace{0.1in}

\noindent {\bf Task 7: Lossy source compression.}   Let $k \geq 0$. There are two parties Alice and Bob. Alice holds a random variable $X$ and Bob holds a random variable $Y$. Alice sends a message to Bob and Bob outputs a random variable $Z$ such that $\Pr \left\{d(X,Z) \geq k\right\} \leq  \eps$, where $d: \cX \times \cZ \to (0,\infty)$ is a {\em distortion} measure. There is no shared randomness allowed between Alice and Bob. 

\vspace{0.1in} 

This problem was studied in the asymptotic i.i.d setting in \cite{WynerZ76}, in the  non-i.i.d. setting in \cite{IwataM02}, in the finite blocklength regime in \cite{KostinaV12, Kostina13} and in the second order setting in \cite{WatanabeKT15, YasAG13}. We show the following achievability result which follows as a corollary of Theorem~\ref{BRachievabilityconverse}. We could alternatively use Theorem~\ref{extensionachievability}, which is also a near optimal characterization of Task $1$.
\begin{corollary}[Achievability for Task 7]
\label{lossy}
Let $\eps, \delta_1, \delta_2 \in (0,1)$. Let $R$ be a natural number such that,
\beq
\label{achievablerd}
R \geq \min_{M,f} \left(\dseps{\eps}(p_{XMY}\| p_Y(p_{X\mid Y} \times p_{M\mid Y})) + 2\log\frac{1}{\delta_2}\right),
\enq
where  $M$ and $f$ satisfy $M-X-Y$  and $\Pr\left\{d(X,f(Y,M))  \geq  k \right\}  \leq \delta_1.$ There exists a protocol with communication $R$ bits from Alice to Bob, in which Bob outputs a random variable $Z$ such that $\Pr \left\{d(X,Z) \geq k\right\} \leq \eps+\delta_1+ 3\delta_2.$ 

Moreover, for any $\delta\in (0,1)$ and for any protocol in which Alice communicates $R$ bits to Bob and Bob outputs a random variable $Z$ such that $\Pr \left\{d(X,Z) \geq k\right\} \leq \eps$, there exists a random variable $M$ and a function $f$ such that $Z= f(Y,M)$ and 
\beq
R \geq \left(\dseps{\eps/(1-\delta)}(p_{XMY}\| p_Y(p_{X\mid Y} \times p_{M\mid Y})) - \log\frac{1}{\delta}\right).
\enq
\end{corollary}

Next we consider the following problem which was first studied by Slepian-Wolf~\cite{SlepianW73} in the asymptotic setting. Its one-shot version was studied in \cite{Warsi2016, uteymatsu-matsuta-2014}. Its second order analysis was given in \cite{TanKosut14}. 

\vspace{0.1in}

\noindent{\bf Task 8: Two-senders-one-receiver source compression.}   There are three parties Alice, Bob and Charlie. Alice possesses a random variable $X,$ Bob possesses a random variable $Y$. Alice and Bob both send a message each to Charlie who at the end outputs random variables $(\hat{X},\hat{Y})$ such that $\Pr\left\{(X,Y) \neq (\hat{X},\hat{Y}) \right\} \leq \eps.$ There is no shared randomness allowed between any parties. 

\vspace{0.1in} 

We show the following achievability result for this task which follows from Theorem~\ref{slepianwolfachieve}. We also provide a matching converse.
\begin{corollary}[Achievability for Task 8]
\label{swach}
Fix $\eps, \delta \in (0,1)$. Let $(X,X)\sim p_{XX}$, where $p_{XX}(x,x)= p_X(x)$ and $(Y,Y)\sim p_{YY}$, where $p_{YY}(y,y)= p_Y(y)$. Let $S,T$ be random variables taking values over the sets $\cX, \cY$ respectively. Let $R_A,R_B, r_A, r_B$ be natural numbers such that,
\begin{align*}
R_A+r_A &\geq \dseps{0}(p_{XX} \| p_X \times p_S) +    2\log \frac{3}{\delta} ,\\
R_B +r_B& \geq \dseps{0}(p_{YY} \| p_T \times p_Y) +    2\log \frac{3}{\delta},
\end{align*}
and
$$\Pr_{(x,y)\leftarrow p_{XY}}\left\{\frac{p_{XY}(x,y)}{p_S(x)p_Y(y)} \leq \frac{2^{r_A}}{\delta} \quad\text{or}\quad \frac{p_{XY}(x,y)}{p_X(x)p_T(y)} \leq \frac{2^{r_B}}{\delta} \quad\text{or}\quad\frac{p_{XY}(x,y)}{p_S(x)p_T(y)} \leq \frac{2^{r_A+r_B}}{\delta}\right\} \leq \eps.$$
In particular, if $S,T$ are distributed according to the uniform probability distribution, it suffices to have $R_A, R_B$ satisfying
\begin{eqnarray}
\label{eq:choiceslepianwolf}
&&\Pr_{(x,y)\leftarrow p_{XY}}\bigg\{\log \frac{1}{p_{X\mid Y=y}(x)} \geq R_A - 3\log\frac{3}{\delta} \quad\text{or}\quad \log\frac{1}{p_{Y\mid X=x}(y)} \geq R_B- 3\log\frac{3}{\delta}\nonumber\\ && \hspace{2cm} \quad\text{or}\quad \log\frac{1}{p_{XY}(x,y)}\geq R_A + R_B - 5\log\frac{3}{\delta}\bigg\} \leq \eps.
\end{eqnarray}
There exists a protocol with communication $R_A$ bits from Alice to Charlie and $R_B$ bits from Bob to Charlie, in which Charlie outputs random variable pair $(\hat{X},\hat{Y})$ such that $\Pr\left\{(X,Y) \neq (\hat{X},\hat{Y}) \right\} \leq \eps + 8\delta.$ 

Furthermore, for any $\delta\in (0,1)$ and any protocol where Alice sends $R_A$ bits to Charlie, Bob sends $R_B$ bits to Charlie and Charlie outputs random variable pair $(\hat{X},\hat{Y})$ such that $\Pr\left\{(X,Y) \neq (\hat{X},\hat{Y}) \right\} \leq \eps$, it holds that
\begin{eqnarray}
&&\Pr_{(x,y,x',y') \drawn p_{XY\hat{X}\hat{Y}}}\bigg\{\log\frac{1}{p_{Y\mid X=x}(y)}> R_A + 2\log\frac{1}{\delta} \text{ or } \log\frac{1}{p_{Y\mid X=x}(y)}> R_B + 2\log\frac{1}{\delta} \nonumber\\ 
&& \hspace{3cm} \text{ or }\log\frac{1}{p_{X,Y}(x,y)}> R_A+R_B+2\log\frac{1}{\delta}\bigg\} \leq \frac{\eps}{1-\delta}+3\delta.
\end{eqnarray}
\end{corollary}

Next we consider the following task which was first studied by Wyner~\cite{Wyner75} in the asymptotic and i.i.d. setting, subsequently in the {\em information-spectrum} setting in~\cite{miyakaye-kanaya-1995, WatanabeKT15, Verdu12}, in the second order setting by \cite{WatanabeKT15, Verdu12} and in the one-shot case in~\cite{Warsi2016, uteymatsu-matsuta-2015}. 

\vspace{0.1in}

\noindent {\bf Task 9: Source compression with coded side information available at the decoder.}  There are three parties Alice, Bob and Charlie. Alice possesses a random variable $X,$ Bob possesses a random variable $Y$. Alice and Bob both send a message each to Charlie who at the end outputs a random variable $\hat{X}$ such that $\Pr\left\{X \neq \hat{X} \right\} \leq \eps.$
 
 \vspace{0.1in} 
 
We show the following achievability result for this task which follows as a corollary from Theorem~\ref{slepianwolfachieve}.

\begin{corollary}[Achievability for Task 9]
\label{helperach}
Let $(X,X)\sim p_{XX}$, where $p_{XX}(x,x)= p_X(x)$.  Let $S,T$ be random variables taking values over the sets $\cX, \cN$ respectively. Let $R_A,R_B, r_A, r_B$ be natural numbers  such that,
\begin{align*}
R_A+r_A &\geq \dseps{0}(p_{XX} \| p_X \times p_S) +    2\log \frac{3}{\delta} ,\\
R_B +r_B& \geq \dseps{\eta}(p_{YN} \| p_Y \times p_T) +    2\log \frac{3}{\delta},
\end{align*}
and
$$\Pr_{(x,n)\leftarrow p_{XN}}\left\{\frac{p_{XN}(x,n)}{p_S(x)p_N(n)} \leq \frac{2^{r_A}}{\delta} \quad\text{or}\quad \frac{p_{XN}(x,n)}{p_X(x)p_T(n)} \leq \frac{2^{r_B}}{\delta} \quad\text{or}\quad\frac{p_{XN}(x,n)}{p_S(x)p_T(n)} \leq \frac{2^{r_A+r_B}}{\delta}\right\} \leq \eps,$$
where $X-Y-N.$  In particular, if $S$ is distributed according to the uniform probability distribution and $p_T= p_N$, then it is sufficient to have $R_A, R_B$ such that
\begin{eqnarray}
\label{eq:choicehelper}
R_B \geq \dseps{\eta}(p_{YN} \| p_Y \times p_N) + 5\log\frac{1}{\delta}, \quad \Pr_{(x,n)\leftarrow p_{XN}}\bigg\{p_{X \mid N=n}(x) \leq \frac{2^{-R_A}}{\delta^3}\bigg\} \leq \eps-\delta.
\end{eqnarray}
There exists a protocol with communication $R_A$ bits from  Alice to Charlie and $R_B$ bits from Bob to Charlie, in which Charlie outputs random variable $\hat{X}$ such that $\Pr\left\{X \neq \hat{X} \right\} \leq \eps+ \eta+8\delta.$ 

Furthermore, for any $\eta, \delta \in (0,1)$ and any protocol where Alice sends $R_A$ bits to Charlie and Bob sends $R_B$ bits to Charlie and Charlie outputs a $\hat{X}$ satisfying $\Pr\left\{X \neq \hat{X} \right\} \leq \eps$ , there exists a random variable $N$ such that
\begin{equation}
R_B \geq \dseps{\eta}(p_{YN} \| p_Y \times p_N) - \log\frac{1}{\eta}, \quad \Pr_{(x, n)\drawn p_{XN}}\left\{p_{X\mid N=n}(x) \leq \delta^2\cdot 2^{-R_A}\right\} \leq \frac{\eps}{1-\delta} + \delta.
\end{equation}
\end{corollary}

\subsection*{Asymptotic and i.i.d. properties}

As discussed earlier, our achievable communication for Task \ref{task1} is optimal in the asymptotic and i.i.d. setting. Using the asymptotic i.i.d. properties of the information spectrum relative entropy and smooth hypothesis testing divergence from \cite{TomHay13}, we are able to establish the \textit{rate regions} for all the remaining tasks. We discuss the rate regions for Task $3$ (which subsumes Task $2$), Task $5$ (which subsumes Task $4$), Task $7$, Task $8$ and Task $9$.
\begin{itemize}
\item {\bf Task $3$:} The rate region is given as (where we use $R^*_A,R^*_B$ to represent the rates)
\begin{align*}
R^*_A &\geq \mutinf{X}{M} - \mutinf{M}{NZ}, \\
R^*_B &\geq \mutinf{Y}{N} - \mutinf{MZ}{N} ,\\
R^*_A+R^*_B &\geq \mutinf{X}{M} + \mutinf{Y}{N} - \trimutinf{Z}{M}{N},
\end{align*}
where $\trimutinf{Z}{M}{N}= \relent{p_{ZMN}}{p_Z\times p_M\times p_N}$ is the tripartite mutual information.

\item {\bf Task $5$:}  The rate region is given as (where we use $R^*_A,R^*_B$ to represent the rates)
\begin{align*}
R^*_A &\geq \mutinf{X}{M} - \mutinf{M}{Y}, \\
R^*_B &\geq \mutinf{X}{N} - \mutinf{N}{Z} ,\\
R^*_A+R^*_B &\geq \trimutinf{X}{M}{N} - \mutinf{M}{Y} - \mutinf{N}{Z}.
\end{align*}
\item {\bf Task $7$:} The achievable rate is 
$$R^* \geq  \min_{M,f} \left(\condmutinf{X}{M}{Y}\right),$$
where $M$ satisfies $M-X-Y$ and $f$ satisfies $\lim_{n\rightarrow \infty}\Pr\left\{d(X^n,f(Y^n,M^n))  \geq  k \right\} = 0$. Here, $Y^n$ is $Y\times Y \times \ldots Y$ ($n$ times) and similarly $M^n$ is $M \times M \times \ldots M$ ($n$ times).

\item {\bf Task $8$:} The rate region is given as (where we use $R^*_A,R^*_B$ to represent the rates)
\begin{align*}
R^*_A &\geq \mathrm{H}(X) - \mutinf{X}{Y} = \mathrm{H}(X\mid Y), \\
R^*_B &\geq \mathrm{H}(Y) - \mutinf{X}{Y} = \mathrm{H}(Y\mid X) ,\\
R^*_A+R^*_B &\geq \mathrm{H}(XY).
\end{align*}
This recovers the rate region obtained by Slepian and Wolf \cite{SlepianW73}.

\item {\bf Task $9$:} The rate region is given as (where we use $R^*_A,R^*_B$ to represent the rates)
\begin{align*}
R^*_A &\geq \mathrm{H}(X) - \mutinf{X}{N} = \mathrm{H}(X\mid N), \\
R^*_B &\geq \mutinf{Y}{N} - \mutinf{X}{N},\\
R^*_A+R^*_B &\geq \mathrm{H}(X) - \mutinf{X}{N} + \mutinf{Y}{N} = \mathrm{H}(X\mid N) + \mutinf{Y}{N}.
\end{align*}
A subset of this rate region is the one obtained in \cite{Wyner75}.
\begin{align*}
R^*_A &\geq \mathrm{H}(X\mid N), \\
R^*_B &\geq \mutinf{Y}{N}.
\end{align*}
Both rate regions match when taken as a union over all $N$ (which satisfy $X-Y-N$), due to the optimality of the latter. However, for a given $N$, our achievability result also implies the result of Slepian and Wolf \cite{SlepianW73} (by setting $N=Y$). 
\end{itemize} 

\subsection*{Organization} In the next section we present a few information theoretic preliminaries. In Section~\ref{sec:proofs} we present proofs of our results. In Section~\ref{sec:classicalcompression}, we consider the question of near optimality of Task \ref{task1}. In Appendix~\ref{sec:deferred} we present some deferred proofs.  
\section{Preliminaries}
\label{sec:prelim}
In this section we set our notations, make the definitions and state the facts that we will need later for our proofs. 

For a natural number $n$, let $[n]$ denote the set $\{1, 2, \ldots, n\}$. Let random variable $X$ take values in a finite set $\cX$ (all sets we consider in this paper are finite).  We let $p_X$ represent the distribution of $X$, that is for each $x \in \cX$ we let $p_X(x):= \Pr(X=x)$. \suppress{Define $\supp(X) := \{ x\in \cX ~| ~p_X(x) > 0 \}$.} Let random variable $Y$ take values in the set $\cY$.  We say $X$ and $Y$ are independent iff for each $x \in \cX, y \in \cY: p_{XY}(x,y) = p_X(x) \cdot p_Y(y)$ and denote $p_X \times p_Y := p_{XY}$. We say random variables $(X,Y,Z)$ form a Markov chain, represented as  $Y-X-Z$, iff for each $x \in \cX$, $Y|(X=x)$ and $Z|(X=x)$ are independent. For an event $E$, its complement is denoted by $\neg E$. We define various information theoretic quantities below. 
\begin{definition}
Let $\eps > 0 $. Let random variables $X$ and $X'$ take values in $\cX$. Define,
\begin{itemize}
\item $\ell_1$ distance:  $\|p_X-p_{X'}\| := \sum_{x}|p_X(x)-p_{X'}(x)|$.
\item Relative entropy: $\mathrm{D}(p_{X} \| p_{X'}) := \sum_{x \in \cX}p_{X}(x) \log \frac{p_{X}(x)}{p_{X'}(x)}$.
\item Max divergence: $\dinfty(p_X\|p_{X'}) := \max_x \log\frac{p_X(x)}{p_{X'}(x)}$.
\item Smooth max divergence: $\dinftyeps(p_X\|p_{X'}):= \min_{\|p_{X''}-p_X\|\leq \eps} \dinfty(p_{X''}\|p_{X'}).$
\item Max information spectrum divergence: $\dseps{\eps}(p_X\|p_{X'}): = \min\left\{a: \Pr _{x \leftarrow p_X}\left\{\frac{p_X(x)}{p_{X'}(x)}> 2^a\right\}<\eps\right\}$.
\item Smooth hypothesis testing divergence: $\dzeroseps{\eps}(p_X\|p_{X'}) : = \mathrm{max}\left\{- \log (\Pr_{p_{X'}}\left\{\cA \right\}) ~|~  \cA \subseteq \cX ,  \Pr_{p_X}\left\{\cA \right\}\geq 1-\eps\right\}$.
\end{itemize}
\end{definition}

We will use the following facts.

\begin{fact}[\cite{AnshuDJ14}]
\label{klprop}
%\mycomment{notation of this fact is not consistent} 
Let $P$ and $Q$ be two distributions over the set $\cX,$ where $P= \sum_{i}{\lambda_i}P_{i}$ is a convex combination of distributions $\{P_i\}_i.$ It holds that,
$$\mathrm{D}(P \| Q)=  \sum_{i} \lambda_{i}\left(\mathrm{D}(P_i \| Q) - \mathrm{D}(P_i \| P)\right).$$
\end{fact}

\begin{fact}[Monotonicity of relative entropy~\cite{CoverT91}]
\label{fact:monorel}
Let $(X,Y,Z)$ be jointly distributed random variables. It holds that,
$$\mathrm{D}(p_{XYZ} \| p_X \times p_Y \times p_Z) \geq \mathrm{D}(p_{XY} \| p_X \times p_Y).$$
\end{fact}

\begin{fact}[Pinsker's inequality~\cite{CoverT91}]
\label{fact:pinsker}
%\mycomment{notation of this fact is not consistent} 
Let $X$ and $X'$ be two random variables over the set $\cX$. It holds that,
$$\| p_X - p_{X'} \| \leq 2 \cdot \sqrt{\mathrm{D}(p_X \| p_{X'})}.$$
\end{fact}

\begin{fact}
\label{closeprob}
Let $X$ and $X'$ be two random variables over the set $\cX$, such that $\frac{1}{2}\|p_X-p_{X'}\|_1 \leq \eps$ for some $\eps\in (0,1)$. For every $\delta\in (0,1)$, it holds that
$$\Pr_{x \drawn p_X}\left\{\frac{p_X(x)}{p_{X'}(x)} \geq \frac{1}{\delta}\right\} \leq \frac{\eps}{1-\delta}.$$
\end{fact}
\begin{proof}
Define the set $$\cA:=\left\{x\in \cX:\frac{p_X(x)}{p_{X'}(x)} \geq \frac{1}{\delta}\right\}.$$
Then 
$$\eps \geq \frac{1}{2}\|p_X-p_{X'}\|_1 \geq \Pr_{p_X}(\cA) - \Pr_{p_{X'}}(\cA) \geq \Pr_{p_X}(\cA) - \delta\Pr_{p_{X}}(\cA) = (1-\delta)\Pr_{p_{X}}(\cA).$$
Thus, 
$$\Pr\{p_X\}\{\cA\}\leq \frac{\eps}{1-\delta} \implies \Pr_{x \drawn p_X}\left\{\frac{p_X(x)}{p_{X'}(x)} \geq \frac{1}{\delta}\right\} \leq \frac{\eps}{1-\delta}.$$
This completes the proof.
\end{proof}
\begin{fact}[Monotonicity under maps~\cite{CoverT91}]
\label{fact:monoprob}
Let $X$  be a random variable  distributed over the set $\cX$. Let $f: \cX \rightarrow \cZ$ be a function. Let random variable $Z$, distributed over $\cZ$ be defined as,
$$ \Pr\{ Z = z\} :=  \frac{\Pr\{X \in f^{-1}(z) \}} {\sum_{z'}  \Pr\{X  \in f^{-1}(z') \}} .$$
Similarly define random variable $Z'$ from random variable $X'$. It holds that,
$$\| p_X  - p_{X'} \| \geq \| p_Z - p_{Z'} \| .$$
\end{fact}

Following convex-split lemma from~\cite{AnshuDJ14} is a main tool that we use. \cite{AnshuDJ14} provided a proof for a quantum version of this lemma and the proof of the classical version that we consider follows on similar lines.  We defer the proof to  Appendix, which uses a procedure from \cite{ABJT18} for perturbing distributions in desired manner without changing one of the marginals. 
\begin{fact}[Convex-split lemma~\cite{AnshuDJ14}]
\label{convexcomb}
Let $\eps, \delta\in (0,1)$. Let $(X,M)$ (jointly distributed over $\cX \times \cM$) and $W$ (distributed over $\cM$) be random variables. Let $R$ be a natural number such that,
$$R \geq  \dseps{\eps}(p_{XM} \| p_X\times p_W) +   2\log\frac{3}{\delta}.$$ 
Let $J$ be uniformly distributed in $[2^R]$ and joint random variables $\left(J,X,M_1, \ldots, M_{2^R}\right)$ be distributed as follows:  
\begin{align*}
&\Pr\left\{(X, M_1, \ldots, M_{2^R}) = (x, m_1, \ldots, m_{2^R}) ~|~ J=j \right\}\\
 &=  p_{XM}(x,m_j)\cdot p_{W}(m_1)\cdots p_{W}(m_{j-1}) \cdot p_{W} (m_{j+1})\cdots p_W(m_{2^R}) .
\end{align*}
Then (below for each $j \in [2^R], p_{W_j} =  p_W$), $$ \frac{1}{2}\|p_{XM_1 \ldots M_{2^R}} - p_X\times p_{W_1} \times\ldots \times  p_{W_{2^R}}\| \leq \eps + \delta .$$
\end{fact}

We need the following extension of this lemma whose quantum version was shown in~\cite{AnshuJW17-a}. The proof of the classical version that we consider follows on similar lines and is deferred to Appendix. We again follow the prescription in \cite{ABJT18} for perturbing the tripartite distribution.
\begin{fact}[Bipartite convex-split lemma]
\label{genconvexcomb}
Let $\eps, \delta\in (0,1)$. Let $(X,M,N)$ (jointly distributed over $\cX\times \cM\times \cN$), $U$ (distributed over $\cM$) and $V$ (distributed over $\cN$) be random variables. Let $R_1, R_2$ be natural numbers such that,
$$\Pr_{(x,m,n)\leftarrow p_{XMN}}\left\{\frac{p_{XM}(x,m)}{p_X(x)p_U(m)} \geq \frac{\delta^2}{24}\cdot 2^{R_1} \quad\text{or}\quad \frac{p_{XN}(x,n)}{p_X(x)p_V(n)} \geq \frac{\delta^2}{24}\cdot 2^{R_2} \quad\text{or}\quad\frac{p_{XMN}(x,m,n)}{p_X(x)p_U(m)p_V(n)} \geq \frac{\delta^2}{24}\cdot 2^{R_1+R_2}\right\} \leq \eps.$$
In particular, the following choice of $R_1, R_2$ suffices, with $\eps_1+\eps_2+\eps_3\leq \eps$ and $\eps_1,\eps_2,\eps_3\in (0,1)$.
\begin{align}
\label{choicebipconv}
R_1 & \geq \dseps{\eps_1}(p_{XM} \| p_X\times p_U) + 2\log\frac{5}{\delta} , \nonumber\\
R_2 & \geq \dseps{\eps_2}(p_{XN} \| p_{X}\times p_V) + 2\log\frac{5}{\delta} ,\nonumber\\
R_1+R_2 & \geq \dseps{\eps_3}(p_{XMN} \| p_X\times p_U \times p_V) + 2\log\frac{5}{\delta}.
\end{align}
Let $J$ be uniformly distributed in $[2^{R_1}]$, $K$ be independent of $J$ and be uniformly distributed in $[2^{R_2}]$ and  joint random variables $\left(J,K,X,M_1,\ldots, M_{2^{R_1}},N_1,\ldots, N_{2^{R_2}}\right)$ be distributed as follows:
\begin{align*}
&\Pr\left\{(X,M_1, \ldots, M_{2^{R_1}}, N_1, \ldots, N_{2^{R_2}}) = (x, m_1, \ldots, m_{2^{R_1}}, n_1, \ldots , n_{2^{R_2}})~|~ J=j, K=k\right\}\\
 &=  p_{XMN}(x,m_j,n_k)\cdot p_{U}(m_1)\cdots p_{U}(m_{j-1}) \cdot p_{U}(m_{j+1})\cdots p_{U}(m_{2^{R_1}}) \cdot \\ 
&\hspace{35mm} p_V(n_1)\cdots p_{V}(n_{k-1}) \cdot p_{V}(n_{k+1})\cdots p_{V}(n_{2^{R_2}}).
\end{align*}
Then (below for each $j \in [2^{R_1}], p_{U_j} = p_U$ and for each $k \in [2^{R_2}], p_{V_k} = p_V$),
 $$ \frac{1}{2}\|p_{XM_1 \ldots M_{2^{R_1}}N_1 \ldots N_{2^{R_2}}} - p_X\times p_{U_1} \times\ldots \times p_{U_{2^{R_1}}}\times p_{V_1} \times\ldots \times p_{V_{2^{R_2}}}\| \leq \eps + \delta.$$
\end{fact}
The other main tool that we use is the position-based decoding from~\cite{AnshuJW17} where a quantum version was shown. The proof of the classical version that we consider follows on similar lines and is deferred to Appendix.
\begin{fact}[Position-based decoding~\cite{AnshuJW17}]
\label{fact:position}
Let $\eps, \delta\in (0,1)$. Let $(Y,M)$ (jointly distributed over $\cY \times \cM$) and $W$ (distributed over $\cM$) be random variables. Let $R$ be a natural number such that,
$$ R \leq \max\left\{ \dzeroseps{\eps}(p_{YM} \| p_Y\times p_W) - \log\frac{1}{\delta} , 0 \right\}.$$ 
Let joint random variables $\left(J,Y,M_1,M_2,\ldots, M_{2^R}\right)$ be distributed as follows. Let $J$ be uniformly distributed in $[2^R]$ and  
\begin{align*}
&\Pr\left\{(Y, M_1, M_2, \ldots, M_{2^R}) = (y, m_1, \ldots, m_{2^R}) ~|~ J=j \right\}\\
 &=  p_{YM}(y,m_j)\cdot p_{W}(m_1)\cdots p_{W}(m_{j-1}) \cdot p_{W} (m_{j+1})\cdots p_W(m_{2^R}) .
\end{align*}
There is a procedure to produce a random variable $J'$ from $(Y, M_1, M_2, \ldots, M_{2^R})$  such that $\Pr\{J \neq J'\} \leq \eps + \delta$ and $\frac{1}{2}\|p_{YM_{J'}} - p_{YM}\|\leq \eps+2\delta$.
\end{fact}
We will also need the following extension of this decoding strategy shown in~\cite{AnshuJW17-a} where a (more general) quantum version was shown. The proof of the classical version that we consider follows on similar lines and is deferred to Appendix.
\begin{fact}[Bipartite position-based decoding~\cite{AnshuJW17-a}]
\label{fact:biposition}
Let $\eps, \delta\in (0,1)$. Let $(Y, M,N)$ (jointly distributed over $\cY\times \cM \times \cN$), $U$ (distributed over $\cM$) and $V$ (distributed over $\cN$). Let $R_1, R_2$ be natural numbers such that,
$$\Pr_{(y,m,n)\leftarrow p_{YMN}}\left\{\frac{p_{YMN}(y,m,n)}{p_U(m)p_{YN}(y,n)} \leq \frac{2^{R_1}}{\delta} \quad\text{or}\quad \frac{p_{YMN}(y,m,n)}{p_{YM}(y,m)p_V(n)} \leq \frac{2^{R_2}}{\delta} \quad\text{or}\quad\frac{p_{YMN}(y,m,n)}{p_Y(y)p_U(m)p_V(n)} \leq \frac{2^{R_1+R_2}}{\delta}\right\} \leq \eps.$$
For instance, the following choice of $R_1, R_2$ suffices, with $\eps_1+\eps_2+\eps_3\leq \eps-3\delta$ and $\eps_1,\eps_2,\eps_3\in (0,1)$. 
\begin{align}
\label{choicebipartite}
R_1 & \leq \max \left\{\dzeroseps{\eps_1}(p_{YMN} \|  p_U \times p_{YN}) - 2\log\frac{1}{\delta}, 0 \right\}\nonumber\\
R_2 & \leq \max \left\{\dzeroseps{\eps_2}(p_{YMN} \|  p_{YM} \times p_V) - 2\log\frac{1}{\delta}, 0 \right\}\nonumber\\
R_1+R_2 & \leq \max \left\{\dzeroseps{\eps_3}(p_{YMN} \| p_Y\times p_U \times p_V) - 2\log\frac{1}{\delta}, 0 \right\}.
\end{align}
Let joint random variables $\left(J,K, Y, M_1,\ldots, M_{2^{R_1}},N_1,\ldots, N_{2^{R_2}}\right)$ be distributed as follows. Let $J$ be uniformly distributed in $[2^{R_1}]$. Let $K$ be independent of $J$ and be uniformly distributed in $[2^{R_2}]$. Let,
\begin{align*}
&\Pr\left\{(Y, M_1 \ldots M_{2^{R_1}} N_1 \ldots N_{2^{R_2}}) = (y, m_1, \ldots, m_{2^{R_1}}, n_1, \ldots , n_{2^{R_2}}) ~|~ J=j, K=k\right\}\\
 &=  p_{YMN}(y, m_j, n_k)\cdot p_{U}(m_1)\cdots p_{U}(m_{j-1}) \cdot p_{U}(m_{j+1})\cdots p_{U}(m_{2^{R_1}}) \cdot \\ 
&\hspace{35mm} p_V(n_1)\cdots p_{V}(n_{k-1}) \cdot p_{V}(n_{k+1})\cdots p_{V}(n_{2^{R_2}}).
\end{align*}
There is a procedure to produce random variables $(J',K')$ from $\left(Y, M_1,\ldots, M_{2^{R_1}},N_1,\ldots, N_{2^{R_2}}\right)$   such that $\Pr\{(J, K) \neq (J', K')\} \leq \eps + 3\delta$ and $\frac{1}{2}\|p_{YM_{J'}N_{K'}} - p_{YMN}\| \leq \eps + 6\delta$.
\end{fact}

\section{Proofs of our results} \label{sec:proofs}
In this section we present proofs of our results mentioned in the Introduction~\ref{intro}.

\begin{proofof}{Theorem~\ref{extensionachievability}} Let $E$ be such that $Y-X-(M,E) $.
Let $R, r$ be natural numbers such that,
\begin{align*}
r & \leq  \max \left\{\dzeroseps{\eps}(p_{Y M E} \| p_{Y} \times p_T) -  \log \frac{1}{\delta} , 0\right\} , \\
R + r & \geq  \dseps{\delta}(p_{X M E} \| p_{X} \times p_T) +   2\log\frac{3}{\delta} .
\end{align*}
Let us divide $[2^{R+r}]$ into $2^R$ subsets, each of size $2^r$. This division is known to both Alice and Bob. For $j \in [2^{R+r}]$, let $\cB(j)$ denote the subset corresponding to $j$. Let us invoke convex-split lemma (Fact~\ref{convexcomb}) with $X \leftarrow X, M  \leftarrow (M, E), W \leftarrow T$ and $ R \leftarrow R+r$  to obtain joint random variables $(J,X, M_1, \ldots, M_{2^{R+r}})$.  Let us first consider a fictitious protocol $\cP'$ as follows. 

\vspace{0.1in}

\noindent {\bf Fictitious protocol $\cP'$:}  Alice possesses random variable $X$, Bob possesses random variable $Y$ and they share $(M_1, \ldots, M_{2^{R+r}})$ as public randomness (from the joint random variables $(X, M_1, \ldots, M_{2^{R+r}})$ above). 

\vspace{0.1in}

\noindent {\bf Alice's operations:} Alice generates $J$ from $(X, M_1, \ldots, M_{2^{R+r}})$, using the conditional distribution of $J$ given $(X, M_1, \ldots, M_{2^{R+r}})$, and communicates $\cB(J)$ to Bob. This can be done using $R$ bits of communication. A similar encoding strategy is used in the works \cite{SongCP15, GoldfeldCP16} (see also the references therein).

\vspace{0.1in}

\noindent {\bf Bob's operations:} Bob performs position-based decoding as in Fact~\ref{fact:position} using $Y$ and the subset $\cB(J)$, by letting $Y \leftarrow Y, M \leftarrow (M, E), W \leftarrow T$ and $R \leftarrow r$, and determines $J'$. Let $(M', E') := M_{J'}$. Bob outputs $M'$. 

\vspace{0.1in}

From Fact~\ref{fact:position} we have 
\begin{align}
\frac{1}{2}\|  p_{X Y M} - p_{X Y M'}   \|  & \leq \eps+2\delta. \label{eq:lowdist}
\end{align}
Now consider the actual protocol $\cP$.

\vspace{0.1in}

\noindent {\bf Actual protocol $\cP$:}  Alice possesses random variable $X$ and Bob possesses random variable $Y$. Alice and Bob share  $2^{R+r}$ i.i.d. copies of the random variable $T$, denoted $\big\{ T_1, T_2, \ldots, T_{2^{R+r}}\big\}$. Alice and Bob proceed as in $\cP'$. Therefore the only difference in $\cP$ and $\cP'$ is shared randomness. Let $\hat{M}$ be the output of Bob in $\cP$. From convex-split lemma (Fact~\ref{convexcomb}), 
$$\frac{1}{2}\|p_{X M_1 \ldots M_{2^{R+r}}}- p_{X}  \times p_{T_1}  \times  \ldots \times  p_{T_{2^{R+r}}}\| \leq 2\delta.$$
Thus, 

\begin{align*}
\frac{1}{2}\|p_{X Y M} - p_{X Y \hat{M}}\| & \overset{a}\leq \frac{1}{2}\|p_{X M_1 \ldots M_{2^{R+r}}}- p_{X}  \times p_{T_1}  \times  \ldots \times  p_{T_{2^{R+r}}}\|+ \frac{1}{2}\|p_{X Y M} - p_{X Y M'}\|   \\
&\overset{b}\leq  2\delta + \eps+ 2\delta \leq \eps+ 4\delta.
\end{align*}
where (a) follows from the property $M-X-Y$  and (b) follows from Equation~\eqref{eq:lowdist}. This shows the desired. 
\end{proofof}

\begin{proofof}{Theorem~\ref{slepianwolfachieve}}
Let $R_A,r_A, R_B,r_B$ be natural numbers that satisfy the constraints in the statement of the theorem, that is,

\begin{align*}
R_A+r_A &\geq \dseps{\eta_1}(p_{XM} \| p_X \times p_S) +    2\log \frac{3}{\delta} ,\\
R_B +r_B& \geq \dseps{\eta_2}(p_{YN} \| p_Y \times p_T) +    2\log \frac{3}{\delta},
\end{align*}
and
$$\Pr_{(m,n)\leftarrow p_{MN}}\left\{\frac{p_{MN}(m,n)}{p_U(m)p_N(n)} \leq \frac{2^{r_A}}{\delta} \quad\text{or}\quad \frac{p_{MN}(m,n)}{p_M(m)p_V(n)} \leq \frac{2^{r_B}}{\delta} \quad\text{or}\quad\frac{p_{MN}(m,n)}{p_U(m)p_V(n)} \leq \frac{2^{r_A+r_B}}{\delta}\right\} \leq \eps.$$

Let us divide $[2^{R_A+r_A}]$ into $2^{R_A}$ subsets, each of size $2^{r_A}$. This division is known to both Alice and Charlie. For $j \in [2^{R_A+r_A}]$, let $\cB(j)$ denote the subset corresponding to $j$. Similarly let us divide $[2^{R_B+r_B}]$ into $2^{R_B}$ subsets, each of size $2^{r_B}$. This division is known to both Bob and Charlie. For $k \in [2^{R_B+r_B}]$, let $\cB(k)$ denote the subset corresponding to $k$.

Let us invoke convex-split lemma (Fact~\ref{convexcomb}) two times with $X \leftarrow X, M  \leftarrow M, W \leftarrow S, R \leftarrow R_A+r_A$ and $X \leftarrow Y, M \leftarrow N, W \leftarrow T,  R \leftarrow R_B+r_B$  to obtain joint random variables $(J,K,X,Y,M_1, \ldots, M_{2^{R_A+r_A}},N_1, \ldots, N_{2^{R_B+r_B}})$. 

Let us first consider a fictitious protocol $\cP'$ as follows.

\vspace{0.1in}

\noindent {\bf Fictitious protocol $\cP'$:}  Let Alice and Charlie share $(M_1, \ldots, M_{2^{R_A+r_A}})$ as public randomness.  Let Bob and Charlie share $(N_1, \ldots, N_{2^{R_B+r_B}})$ as public randomness. 

\vspace{0.1in}

\noindent {\bf Alice's operations:} Alice generates $J$ from $(X,M_1, \ldots, M_{2^{R_A+r_A}})$, using the conditional distribution of $J$ given $(X,M_1, \ldots, M_{2^{R_A+r_A}})$, and communicates $\cB(J)$ to Charlie. This can be done using $R_A$ bits of communication. 

\vspace{0.1in}

\noindent {\bf Bob's operations:} Bob generates $K$ from $(Y,N_1, \ldots, N_{2^{R_B+r_B}})$, using the conditional distribution of $K$ given $(Y,N_1, \ldots, N_{2^{R_B+r_B}})$, and communicates $\cB(K)$ to Charlie. This can be done using $R_B$ bits of communication. 

\vspace{0.1in}

\noindent {\bf Charlie's operations:} Charlie performs bipartite position-based decoding as in Fact~\ref{fact:biposition} inside the subset $\cB(J) \times \cB(K)$, by letting $Y \leftarrow \phi, M \leftarrow M, N \leftarrow N, U\leftarrow S, V\leftarrow T, R_A \leftarrow r_A $ and $R_B \leftarrow r_B$ (where $\phi$ is empty assignment), and determines $(J' , K')$. Charlie outputs $(M', N') := (M_{J'}, N_{K'})$. 

\vspace{0.1in}

Note that Alice and Bob's operation produce the right joint distribution $(J,K,X,Y,M_1, \ldots, M_{2^{R_A+r_A}},N_1, \ldots, N_{2^{R_B+r_B}})$ since $M - X - Y - N$. Therefore from  Fact~\ref{fact:biposition} we have, 
\begin{equation} 
\frac{1}{2}\|p_{XYMN}-p_{XY M' N'}\|  \leq  \eps +6\delta. \label{eq:lowdistslepian}
 \end{equation}

\vspace{0.1in}

Now consider the actual protocol $\cP$.

\vspace{0.1in}

\noindent {\bf Actual protocol $\cP$:}  Alice and Charlie share  $2^{R_A+r_A}$ i.i.d. copies of the random variable $S$, denoted $\big\{ S_1, S_2, \ldots, S_{2^{R_A+r_A}}\big\}$. Bob and Charlie share  $2^{R_B+r_B}$ i.i.d. copies of the random variable $T$, denoted $\big\{ T_1, T_2, \ldots, T_{2^{R_B+r_B}}\big\}$. Alice, Bob and Charlie proceed as in $\cP'$. Therefore the only difference in $\cP$ and $\cP'$ is shared randomness. Let $(\hat{M}, \hat{N})$ represent Charlie's outputs in $\cP$.

\vspace{0.1in}

\noindent From convex-split lemma (Fact~\ref{convexcomb})
$$ \frac{1}{2}\|p_{XYM_1 \ldots M_{2^{R_A+r_A}}N_1 \ldots N_{2^{R_B+r_B}}} - p_{XY} \times p_{S_1}  \times  \ldots \times  p_{S_{2^{R_A+r_A}}} \times  p_{T_1}  \times  \ldots \times  p_{T_{2^{R_B+r_B}}} \| \leq \eta_1+\eta_2+ 2\delta.$$ 
From Fact~\ref{fact:monoprob}, triangle inequality for $\ell_1$ norm and Equation~\eqref{eq:lowdistslepian} we have,
$$  \frac{1}{2}\|p_{XYMN}-p_{XY\hat{M}\hat{N}}\| \leq  \eps+\eta_1+\eta_2+6\delta + 2\delta \leq \eps+ \eta_1+\eta_2+ 8\delta.$$
This establishes the correctness of the protocol. To show that the choices of $R_A, R_B$ in Equation~\eqref{choicetask2} suffices, 
we appeal to the Fourier-Motzkin elimination technique~\cite[Appendix D]{GamalK12} and consider natural numbers $r_A, r_B$ such that 
\begin{align*}
R_A+r_A &\geq \dseps{\eta}(p_{XM} \| p_X \times p_S) +    2\log \frac{3}{\delta} ,\\
R_B +r_B& \geq \dseps{\eta}(p_{YN} \| p_Y \times p_T) +    2\log \frac{3}{\delta} ,\\
r_A & \leq \max\{\dzeroseps{\eps_1}(p_{MN}\| p_S\times p_N)  -  2\log \frac{1}{\delta}, 0\},\\
r_B & \leq \max\{\dzeroseps{\eps_2}(p_{MN}\| p_M\times p_T)  -  2\log \frac{1}{\delta}, 0\},\\
r_A+r_B & \leq \max\{\dzeroseps{\eps_3}(p_{MN}\| p_S\times p_T)  - 2\log \frac{1}{\delta}, 0\}.
\end{align*}
From Equation~\eqref{choicebipartite} in Fact \ref{fact:biposition}, this choice satisfies the constraints of the theorem. This completes the proof. 
\end{proofof}

\begin{proofof}{Theorem~\ref{swsideinf}}  The proof follows on similar lines as the proof of Theorem~\ref{slepianwolfachieve} and we provide a proof sketch here. Let $(R_A,R_B,r_A,r_B)$ be natural numbers such that  (existence  of these numbers is guaranteed by the Fourier-Motzkin elimination technique~\cite[Appendix D]{GamalK12} and the constraints in the statement of the Theorem), 
\begin{align*}
R_A+r_A &\geq \dseps{\delta}(p_{XM} \| p_X \times p_S)  + 2\log \frac{3}{\delta} ,\\
R_B +r_B& \geq \dseps{\delta}(p_{YN} \| p_Y \times p_T)  + 2\log \frac{3}{\delta},\\
r_A & \leq \max \left\{\dzeroseps{\eps_1}(p_{MNZ}\| p_S\times p_{NZ})  - 2\log \frac{1}{\delta} , 0\right\} ,\\
r_B & \leq \max \left\{\dzeroseps{\eps_2}(p_{MZN}\| p_{MZ}\times p_{T})  - 2\log \frac{1}{\delta} , 0\right\},\\
r_A+r_B & \leq \max \left\{\dzeroseps{\eps_3}(p_{MNZ}\| p_S\times p_T \times p_Z)  - 2\log \frac{1}{\delta}, 0\right\} .
\end{align*} 
Let $\cA_1, \cA_2, \cA_3 \subseteq \cM \times \cN \times \cZ$ be such that $\Pr_{p_{MNZ}}\{ \cA_i\} \geq 1 - \eps_i$ for all $i \in \{1, 2, 3\}$ and  
\begin{align*}
\dzeroseps{\eps_1}(p_{MNZ}\| p_S\times p_{NZ}) &= - \log \Pr_{p_S\times p_{NZ}} \left\{\cA_1\right\};\\
\dzeroseps{\eps_2}(p_{MZN}\| p_{MZ}\times p_{T})& = - \log \Pr_{p_{MZ}\times p_{T}} \left\{\cA_2\right\};\\
 \dzeroseps{\eps_3}(p_{MNZ}\| p_S\times p_T \times p_Z) & - \log \Pr_{p_S\times p_{T} \times p_{Z}} \left\{\cA_3\right\}.
\end{align*}
Define $\cA := \cA_1 \cap \cA_2 \cap \cA_3.$

\vspace{0.1in}

\noindent {\bf Protocol $\cP$:}  Shared randomness and Alice and Bob's operations remain same as in the actual protocol $\cP$ of the proof of Theorem~\ref{slepianwolfachieve}. 

\vspace{0.1in}

\noindent {\bf{Charlie's operations:}} Charlie on receiving $\cB(J)$ and $\cB(K)$ from Alice and Bob respectively, performs  bipartite position-based decoding (Fact~\ref{fact:biposition})  by letting $Y \leftarrow Z, M \leftarrow M, N \leftarrow N, U\leftarrow S, V\leftarrow T, R_A \leftarrow r_A $ and $R_B \leftarrow r_B$ and finds $(J', K')$.

\vspace{0.1in}

From Fact~\ref{fact:biposition} (bipartite position-based decoding),  Fact~\ref{fact:monoprob}, the convex-split lemma (Fact \ref{convexcomb}) and  triangle inequality for $\ell_1$ norm it can be argued that 
$ \frac{1}{2}\| p_{XY M N}   -p_{XY \hat{M} \hat{N}} \| \leq \eps_1+\eps_2+\eps_3+ 13\delta.$
\end{proofof}

\begin{proofof}{Theorem~\ref{networkprob}} Let us invoke bipartite convex-split lemma (Fact~\ref{genconvexcomb}) with $X \leftarrow X, M  \leftarrow M, N  \leftarrow N, U \leftarrow S,  V \leftarrow T, R_1 \leftarrow R_B$ and $ R_2 \leftarrow R_C$  to obtain joint random variables $(J,K,X,M_1, \ldots, M_{2^{R_B}},N_1, \ldots, N_{2^{R_C}})$. 

Let us first consider a fictitious protocol $\cP'$ as follows.

\vspace{0.1in}

\noindent {\bf Fictitious protocol $\cP'$:}  Let Alice and Bob share $(M_1, \ldots, M_{2^{R_B}})$ as public randomness.  Let Alice and Charlie share $(N_1, \ldots, N_{2^{R_C}})$ as public randomness. 

\vspace{0.1in}

\noindent {\bf Alice's operations:} Alice generates $(J, K)$ from $(X,M_1, \ldots, M_{2^{R_B}},N_1, \ldots, N_{2^{R_C}})$, using the conditional distribution of $(J, K)$ given $(X,M_1, \ldots, M_{2^{R_B}},N_1, \ldots, N_{2^{R_C}})$. She communicates $J$ to Bob (using $R_B$ bits) and $K$ to Charlie (using $R_C$ bits). 

\vspace{0.1in}

\noindent {\bf Bob's operations:} Bob outputs $M' := M_{J}$. 

\vspace{0.1in}

\noindent {\bf Charlie's operations:}  Charlie outputs $N' := N_{K}$. 
\vspace{0.1in}

\noindent It holds that $p_{XM'N'} = p_{XMN}$. Now consider the actual protocol $\cP$.

\vspace{0.1in}

\noindent {\bf Actual protocol $\cP$:}  Alice and Bob share  $2^{R_B}$ i.i.d. copies of the random variable $S$, denoted $\big\{ S_1, S_2, \ldots, S_{2^{R_B}}\big\}$. Alice and Charlie  share  $2^{R_C}$ i.i.d. copies of the random variable $T$, denoted $\big\{ T_1, T_2, \ldots, T_{2^{R_C}}\big\}$. Alice, Bob and Charlie proceed as in $\cP'$. Therefore the only difference in $\cP$ and $\cP'$ is shared randomness. Let $(\hat{M}, \hat{N})$ represent Bob and Charlie's outputs respectively in $\cP$.

\vspace{0.1in}

From bipartite convex-split lemma (Fact~\ref{genconvexcomb}),
\begin{equation} \label{eq:lowdistsleprev} 
\frac{1}{2}\|p_{XM_1 \ldots M_{2^{R_B}}N_1 \ldots N_{2^{R_C}}}- p_{X}   \times p_{M_1}  \times  \ldots \times  p_{M_{2^{R_B}}} \times  p_{N_1}  \times  \ldots \times  p_{N_{2^{R_C}}} \| \leq \eps + \delta. \end{equation}
From Fact~\ref{fact:monoprob}, triangle inequality for $\ell_1$ norm and Equation~\eqref{eq:lowdistsleprev} we have,
$$  \|p_{XMN}-p_{X\hat{M}\hat{N}}\| \leq   \eps + \delta.$$

\end{proofof}

\begin{proofof}{Theorem~\ref{networkprobs}} The proof follows on similar lines as the proof of Theorem~\ref{networkprob} and we provide a proof sketch here. Let $(R_B, R_C, r_B, r_C)$ be natural numbers such that, 
\begin{align*}
R_B +r_b & \geq \dseps{\eps_1}(p_{XM} \| p_X\times p_S)  + 2\log\frac{5}{\delta_2}, \\
r_b & \leq  \max\left\{\dzeroseps{\delta_1}(p_{MY}\| p_{S}\times p_{Y}) - \log\frac{1}{\delta_2} , 0 \right\},\\
R_C+r_c & \geq \dseps{\eps_2}(p_{XN} \| p_X\times p_T)  + 2\log\frac{5}{\delta_2} ,\\
r_c & \leq  \max\left\{\dzeroseps{\delta_1}(p_{NZ}\| p_{T}\times p_{Z}) - \log\frac{1}{\delta_2} , 0 \right\},\\
R_B+R_C +r_b+r_c & \geq \dseps{\eps_3}(p_{XMN} \| p_{X} \times p_S \times p_T)+ 2\log\frac{5}{\delta_2} .
\end{align*} 
Let $\cA_1 \subseteq \cY \times \cM$ and $\cA_2 \subseteq \cZ \times \cN$ be such that $\Pr_{p_{YM}}\{ \cA_1\} \geq 1 - \delta_1$  and   $\Pr_{p_{ZN}}\{ \cA_2\} \geq 1 - \delta_1$ and,
\begin{align*}
\dzeroseps{\delta_1}(p_{MY}\| p_S\times p_{Y}) &= - \log \Pr_{p_S\times p_{Y}} \left\{\cA_1\right\} ,\\
\dzeroseps{\delta_1}(p_{NZ}\| p_{T}\times p_{Z})& = - \log \Pr_{p_{T}\times p_{Z}} \left\{\cA_2\right\}.
\end{align*}
Let us divide $[2^{R_B + r_B}]$ into $2^{R_B}$ subsets, each of size $2^{r_B}$. This division is known to both Alice and Bob. For $j \in [2^{R_B+r_B}]$, let $\cB(j)$ denote the subset corresponding to $j$. Similarly let us divide $[2^{R_C+r_C}]$ into $2^{R_C}$ subsets, each of size $2^{r_C}$. This division is known to both Alice and Charlie. For $k \in [2^{R_C+r_C}]$, let $\cB(k)$ denote the subset corresponding to $k$.
\vspace{0.1in}

\noindent {\bf Protocol $\cP$:}  Alice and Bob share  $2^{R_B+r_b}$ i.i.d. copies of the random variable $S$, denoted $\big\{ S_1, S_2, \ldots, S_{2^{R_B+r_b}}\big\}$. Alice and Charlie  share  $2^{R_C+r_c}$ i.i.d. copies of the random variable $T$, denoted $\big\{ T_1, T_2, \ldots, T_{2^{R_C+r_c}}\big\}$. 

\vspace{0.1in}

\noindent {\bf Alice's operations:} Alice generates $(J, K)$ as in protocol $\cP$ in the proof of Theorem~\ref{networkprob}. She communicates $\cB(J)$ to Bob (using $R_B$ bits) and $\cB(K)$ to Charlie (using $R_C$ bits). 

\vspace{0.1in}

\noindent {\bf Bob's operations:} Bob performs position-based decoding as in Fact~\ref{fact:position}, by letting $Y \leftarrow Y, M \leftarrow M $, $W\leftarrow S$ and $R \leftarrow R_B $  and determines $J'$. Bob outputs $\hat{M} := M_{J'}$. 

\vspace{0.1in}

\noindent {\bf Charlie's operations:}  Charlie performs position-based decoding as in Fact~\ref{fact:position}, by letting $Y \leftarrow Z, M \leftarrow N $, $W\leftarrow T$ and $R \leftarrow R_C $  and determines $K'$. Charlie outputs $\hat{N} := N_{K'}$. 

\vspace{0.1in}

Using the Fact~\ref{fact:position} (for position-based decoding), Fact~\ref{fact:monoprob}, bipartite convex-split lemma (Fact \ref{genconvexcomb}) and  triangle inequality for $\ell_1$ norm it can be argued that  
$ \| p_{XYZ M N}   -p_{XYZ \hat{M} \hat{N}} \|  \leq \eps + 2\delta_1+5\delta_2  .$

\end{proofof}

\begin{proofof}{Corollary \ref{lossy}} 
We divide the proof in two parts.

\vspace{0.1in}

\noindent {\bf Achievability:} Let $M$ and $f$ be such that they achieve the minimum in Equation~\eqref{achievablerd}. Alice and Bob employ the protocol from Theorem~\ref{BRachievabilityconverse} in which Alice send $R$ bits to Bob and at the end Bob is able to generate $\hat{M}$ such that $\frac{1}{2}\|p_{XYM}-p_{XY\hat{M}}\| \leq \eps+3\delta_2$. Bob then outputs $Z = f(Y,\hat{M}).$ Consider, 
\begin{align*}
\Pr\left\{d(X,f(Y,\hat{M}))  \geq k \right\} & \leq \Pr\left\{d(X,f(Y,M))  \geq  k \right\} +  \frac{1}{2}\|p_{XYM}-p_{XY\hat{M}}\| \\
& \leq \delta_1+  \eps+3\delta_2 .
\end{align*}
This protocol uses shared randomness between Alice and Bob and $\Pr\left\{d(X,f(Y, \hat{M}))  \geq k \right\} \leq   \delta_1 +  \eps+3\delta_2$ averaged over the shared randomness. Hence there exists a fixed shared string between Alice and Bob, conditioned on which $\Pr\left\{d(X,f(Y,\hat{M}))  \geq k \right\} \leq   \delta_1 +  \eps+3\delta_2$ . Fixing this string finally gives us the desired protocol which does not use shared randomness.

\vspace{0.1in}

\noindent {\bf Converse:} For the converse, we take $M$ as Alice's message and $f$ as the function used by Bob for decoding a $Z$. The bound on the number of bits of $M$ now follows from the converse part of Theorem~\ref{BRachievabilityconverse}.
\end{proofof}

\begin{proofof}{Corollary \ref{swach}} We divide the proof in two parts.

\vspace{0.1in}

\noindent {\bf Achievability:} Alice, Bob and Charlie use the protocol in Theorem~\ref{slepianwolfachieve} where we set $M \leftarrow X$ and $N \leftarrow Y$. Let  $(\hat{X}, \hat{Y})$ be the output of Charlie. We have, $\frac{1}{2}\| p_{XYXY} - p_{XY\hat{X} \hat{Y}} \|\leq \eps + 16\delta$ which implies $\Pr\left\{(X,Y) \neq (\hat{X},\hat{Y}) \right\} \leq \eps + 16\delta$. This protocol uses shared randomness between Alice and Bob and $\Pr\left\{(X,Y) \neq (\hat{X},\hat{Y}) \right\} \leq \eps + 16\delta$  averaged over the shared randomness. Hence there exists a fixed shared string conditioned on which $\Pr\left\{(X,Y) \neq (\hat{X},\hat{Y}) \right\} \leq \eps + 16\delta$. Fixing this string gives us the desired protocol which does not use shared randomness.

Suppose $S,T$ are distributed according to the uniform probability distribution. To show that the choice of $R_A, R_B$ as given in Equation~\eqref{eq:choiceslepianwolf} suffices, observe that $\dseps{0}(p_{XX} \| p_X \times p_S) = \log|\cX|$ and $\dseps{0}(p_{YY} \| p_T \times p_Y) = \log|\cY|$. Then 
$$R_A+r_A \geq \log|\cX| +    2\log \frac{3}{\delta} , \quad R_B +r_B \geq \log|\cY| +  2\log \frac{3}{\delta}.$$
This implies that 
\begin{eqnarray*}
&&\Pr_{(x,y)\leftarrow p_{XY}}\left\{\frac{p_{XY}(x,y)}{p_S(x)p_Y(y)} \leq \frac{2^{r_A}}{\delta} \quad\text{or}\quad \frac{p_{XY}(x,y)}{p_X(x)p_T(y)} \leq \frac{2^{r_B}}{\delta} \quad\text{or}\quad\frac{p_{XY}(x,y)}{p_S(x)p_T(y)} \leq \frac{2^{r_A+r_B}}{\delta}\right\} \\ 
&&\leq \Pr_{(x,y)\leftarrow p_{XY}}\bigg\{\frac{p_{XY}(x,y)}{p_S(x)p_Y(y)} \leq \frac{9\cdot 2^{-R_A}|\cX|}{\delta^3} \quad\text{or}\quad \frac{p_{XY}(x,y)}{p_X(x)p_T(y)} \leq \frac{9\cdot 2^{-R_B}|\cY|}{\delta^3}\\ && \hspace{2cm} \quad\text{or}\quad\frac{p_{XY}(x,y)}{p_S(x)p_T(y)} \leq \frac{81\cdot 2^{-R_A-R_B}|\cX||\cY|}{\delta^5}\bigg\}\\
&&= \Pr_{(x,y)\leftarrow p_{XY}}\bigg\{\frac{p_{XY}(x,y)|\cX|}{p_Y(y)} \leq \frac{9\cdot 2^{-R_A}|\cX|}{\delta^3} \quad\text{or}\quad \frac{p_{XY}(x,y)|\cY|}{p_X(x)} \leq \frac{9\cdot 2^{-R_B}|\cY|}{\delta^3}\\ && \hspace{2cm} \quad\text{or}\quad p_{XY}(x,y)|\cX||\cY| \leq \frac{81\cdot 2^{-R_A-R_B}|\cX||\cY|}{\delta^5}\bigg\}\\
&&= \Pr_{(x,y)\leftarrow p_{XY}}\bigg\{p_{X\mid Y=y}(x) \leq \frac{9\cdot 2^{-R_A}}{\delta^3} \quad\text{or}\quad p_{Y\mid X=x}(y) \leq \frac{9\cdot 2^{-R_B}}{\delta^3}\quad\text{or}\quad p_{XY}(x,y)\leq \frac{81\cdot 2^{-R_A-R_B}}{\delta^5}\bigg\}\\
&&= \Pr_{(x,y)\leftarrow p_{XY}}\bigg\{\log \frac{1}{p_{X\mid Y=y}(x)} \geq R_A - 3\log\frac{3}{\delta} \quad\text{or}\quad \log\frac{1}{p_{Y\mid X=x}(y)} \geq R_B- 3\log\frac{3}{\delta}\\ && \hspace{2cm} \quad\text{or}\quad \log\frac{1}{p_{XY}(x,y)}\geq R_A + R_B - 5\log\frac{3}{\delta}\bigg\}\\ && \leq \eps.
\end{eqnarray*}
This completes the achievability proof.

\vspace{0.1in}

\noindent {\bf Converse:} For the converse, let $M$ be the message from Alice of $R_A$ bits and $N$ be the message from Bob of $R_B$ bits. It holds that $M-X-Y-N$. Let $U, V$ be uniform distributions over Alice and Bob's messages respectively. Let the output by Charlie be $(\hat{X},\hat{Y})$ which satisfies $\Pr\{(\hat{X},\hat{Y}) \neq (X,Y)\}\leq \eps$. Define random variables $(X^*,Y^*)$ jointly correlated with $(X,Y)$ as follows.
$$p_{X^*Y^*\mid X=x, Y=y}(x', y') = 1 \text{ iff } x=x' \text{ and } y=y'.$$
It holds that
\begin{eqnarray}
\label{errguarantee}
&&\frac{1}{2}\|p_{XY\hat{X}\hat{Y}}-p_{XYX^*Y^*}\|= \frac{1}{2}\sum_{x,y}p_{XY}(x,y)\|p_{\hat{X}\hat{Y}\mid X=x,Y=y} - p_{X^*Y^*\mid X=x,Y=y}\| \nonumber\\
&&= \frac{1}{2}\sum_{x,y}p_{XY}(x,y)\left(1-p_{\hat{X}\hat{Y}\mid X=x,Y=y}(x,y)+\sum_{x'\neq x \text{ or }y'\neq y}p_{\hat{X}\hat{Y}\mid X=x,Y=y}(x',y') \right)\nonumber\\
&& = \sum_{x,y}p_{XY}(x,y)\left(\sum_{x'\neq x \text{ or }y'\neq y}p_{\hat{X}\hat{Y}\mid X=x,Y=y}(x',y') \right)\nonumber\\
&&= \Pr\{(\hat{X},\hat{Y}) \neq (X,Y)\}\leq \eps.
\end{eqnarray}
Consider the following set of inequalities.
\begin{eqnarray}
\label{convbaseid}
p_{XM\mid Y=y, N=n}(x,m) &=& p_{XM\mid Y=y}(x,m) \leq 2^{R_A}p_{X\mid Y=y}(x)p_U(m), \nonumber \\
p_{YN\mid X=x, M=m}(y,n) &=& p_{YN\mid X=x}(y,n) \leq 2^{R_A}p_{Y\mid X=x}(y)p_V(n)\nonumber\\
p_{M,N\mid X=x,Y=y}(m,n) &\leq& 2^{R_A+R_B}p_U(m)p_V(n).
\end{eqnarray}
By data-processing (for Charlie's operation) on the first inequality in Equation~\eqref{convbaseid}, we now conclude that
\begin{eqnarray*}
p_{X\hat{X}\hat{Y}\mid Y=y}(x, x',y') = p_{X\hat{X}\hat{Y}\mid Y=y, N=n}(x,x',y')\leq 2^{R_A}p_{X\mid Y=y}(x)p_{W_1\mid Y=y}(x',y'),
\end{eqnarray*}
where $p_{W_1\mid Y=y}$ is some distribution. Similarly, second and third inequalities of Equation~\eqref{convbaseid} give the inequalities
\begin{eqnarray*}
p_{Y\hat{X}\hat{Y}\mid X=x}(y, x',y') \leq 2^{R_B}p_{Y\mid X=x}(y)p_{W_2\mid X=x}(x',y'), \quad p_{\hat{X}\hat{Y}\mid X=x,Y=y}(x',y') \leq 2^{R_A+R_B} p_{W_3}(x',y'),
\end{eqnarray*}
for some distributions $(W_2\mid X=x)$ and $W_3$. Collectively and rearranging, we conclude
\begin{eqnarray}
\label{rearrid}
p_{X\hat{X}\hat{Y}Y}(x, x',y, y') &\leq& 2^{R_A}p_{X\mid Y=y}(x)p_{W_1Y}(x',y', y), \nonumber \\
p_{X\hat{X}Y\hat{Y}}(x,x', y,y') &\leq& 2^{R_B}p_{Y\mid X=x}(y)p_{W_2 X}(x',y',x)\nonumber\\
p_{X\hat{X}Y\hat{Y}}(x,x',y,y') &\leq& 2^{R_A+R_B}p_{XY}(x,y)p_{W_3}(x',y')
\end{eqnarray}
From Fact \ref{closeprob}, it holds that
$$\Pr_{(x,x',y,y') \drawn p_{XX^*YY^*}}\left\{\frac{p_{XX^*YY^*}(x,x',y,y')}{p_{X\hat{X}Y\hat{Y}}(x,x',y,y')} \geq \frac{1}{\delta}\right\} \leq \frac{\eps}{1-\delta}.$$
Combining with Equation \ref{rearrid}, we conclude
\begin{eqnarray*}
&&\Pr_{\overset{(x,x',y,y') \drawn}{p_{XX^*YY^*}}}\bigg\{\frac{p_{XX^*YY^*}(x,x',y,y')}{p_{X\mid Y=y}(x)p_{W_1Y}(x',y', y)} \geq \frac{2^{R_A}}{\delta} \text{ or } \frac{p_{XX^*YY^*}(x,x',y,y')}{p_{Y\mid X=x}(y)p_{W_2 X}(x',y',x)} \geq \frac{2^{R_B}}{\delta}\\
\hspace{3cm}&&\text{ or } \frac{p_{XX^*YY^*}(x,x',y,y')}{p_{XY}(x,y)p_{W_3}(x',y')} \geq \frac{2^{R_A+R_B}}{\delta}\bigg\} \leq \frac{\eps}{1-\delta}.
\end{eqnarray*}
Since $x=x'$ and $y=y'$ for all $(x,x',y,y')\drawn p_{XX^*YY^*}$, above can be rewritten as
\begin{eqnarray}
\label{eq:XYid}
&&\Pr_{\overset{(x,y) \drawn}{p_{XY}}}\bigg\{\frac{p_{XY}(x,y)}{p_{X\mid Y=y}(x)p_{W_1Y}(x,y, y)} \geq \frac{2^{R_A}}{\delta} \text{ or } \frac{p_{XY}(x,y)}{p_{Y\mid X=x}(y)p_{W_2 X}(x,y,x)} \geq \frac{2^{R_B}}{\delta}\nonumber\\
\hspace{3cm}&&\text{ or } \frac{p_{XY}(x,y)}{p_{XY}(x,y)p_{W_3}(x,y)} \geq \frac{2^{R_A+R_B}}{\delta}\bigg\} \leq \frac{\eps}{1-\delta}.
\end{eqnarray}
Define
$$\cB:=\{(x,y): p_{W_1Y}(x,y, y)> \frac{1}{\delta}p_{XY}(x,y)\}.$$
Observe that
$$\Pr_{(x,y)\drawn p_{XY}}\left\{\cB\right\}= \sum_{(x,y)\in \cB}p_{XY}(x,y) \leq \delta \sum_{(x,y)\in \cB}p_{W_1Y}(x,y,y)\leq \delta \sum_{x,y',y}p_{W_1Y}(x,y',y) \leq \delta.$$
In a similar fashion, we conclude that
$$\Pr_{(x,y) \drawn p_{XY}}\left\{\frac{p_{W_1Y}(x,y,y)}{p_{XY}(x,y)}> \frac{1}{\delta} \text{ or } \frac{p_{W_2 X}(x,y,x)}{p_{XY}(x,y)}> \frac{1}{\delta} \text{ or } \frac{p_{W_3}(x,y)}{p_{XY}(x,y)}> \frac{1}{\delta}\right\} \leq 3\delta.$$
Along with Equation \ref{eq:XYid}, this implies
\begin{eqnarray*}
&&\Pr_{(x,y) \drawn p_{XY}}\bigg\{\frac{1}{p_{X\mid Y=y}(x)} \geq \frac{2^{R_A}}{\delta^2} \text{ or } \frac{1}{p_{Y\mid X=x}(y)} \geq \frac{2^{R_B}}{\delta^2} \text{ or }\frac{1}{p_{XY}(x,y)} \geq \frac{2^{R_A+R_B}}{\delta^2}\bigg\} \leq \frac{\eps}{1-\delta} + 3\delta.
\end{eqnarray*}
This completes the proof.
\end{proofof}

\begin{proofof}{Corollary \ref{helperach}} We divide the proof in two parts.

\vspace{0.1in}

\noindent {\bf Achievability:} Alice, Bob and Charlie use the protocol in Theorem~\ref{slepianwolfachieve} where we set $M  \leftarrow  X$ and $N  \leftarrow N$. Let  $(\hat{X}, \hat{N})$ be the output of Charlie. We have, $\frac{1}{2}\| p_{XYXN} - p_{XY\hat{X} \hat{N}} \|\leq \eps+\eta+8\delta$ which implies $\Pr\left\{X \neq \hat{X} \right\} \leq \eps+\eta+8\delta$. This protocol uses shared randomness between Alice and Bob and $\Pr\left\{X \neq \hat{X} \right\} \leq  \eps+\eta+8\delta$  averaged over the shared randomness. Hence there exists a fixed shared string conditioned on which $\Pr\left\{X \neq \hat{X} \right\} \leq  \eps+\eta+8\delta$. Fixing this string gives us the desired protocol which does not use shared randomness.

Suppose $S$ is distributed according to the uniform probability distribution and $p_T= p_N$. To show that the choice of $R_A, R_B$ as given in Equation~\eqref{eq:choicehelper} suffices, observe that $\dseps{0}(p_{XX} \| p_X \times p_S) = \log|\cX|$. Then 
$$R_A+r_A \geq \log|\cX| +  2\log \frac{3}{\delta} , \quad R_B +r_B \geq \dseps{\eta}(p_{YN} \| p_Y \times p_N) +  2\log \frac{3}{\delta}.$$
This implies that
\begin{eqnarray}
\label{eq:helperinequalities}
&&\Pr_{(x,n)\leftarrow p_{XN}}\left\{\frac{p_{XN}(x,n)}{p_S(x)p_N(n)} \leq \frac{2^{r_A}}{\delta} \quad\text{or}\quad \frac{p_{XN}(x,n)}{p_X(x)p_N(n)} \leq \frac{2^{r_B}}{\delta} \quad\text{or}\quad\frac{p_{XN}(x,n)}{p_S(x)p_N(n)} \leq \frac{2^{r_A+r_B}}{\delta}\right\}\nonumber\\
&&\leq \Pr_{(x,n)\leftarrow p_{XN}}\bigg\{\frac{p_{XN}(x,n)}{p_S(x)p_N(n)} \leq \frac{9\cdot 2^{-R_A}|\cX|}{\delta^3} \quad\text{or}\quad \frac{p_{XN}(x,n)}{p_X(x)p_N(n)} \leq \frac{9\cdot 2^{-R_B + \dseps{\eta}(p_{YN} \| p_Y \times p_N)}}{\delta^3} \nonumber\\ && \hspace{2cm} \quad\text{or}\quad\frac{p_{XN}(x,n)}{p_S(x)p_N(n)} \leq \frac{81\cdot 2^{-R_A-R_B + \dseps{\eta}(p_{YN} \| p_Y \times p_N)}|\cX|}{\delta^5}\bigg\}\nonumber\\
&&= \Pr_{(x,n)\leftarrow p_{XN}}\bigg\{p_{X \mid N=n}(x) \leq \frac{9\cdot 2^{-R_A}}{\delta^3} \quad\text{or}\quad \frac{p_{XN}(x,n)}{p_X(x)p_N(n)} \leq \frac{9\cdot 2^{-R_B + \dseps{\eta}(p_{YN} \| p_Y \times p_N)}}{\delta^3} \nonumber\\ && \hspace{2cm} \quad\text{or}\quad p_{X \mid N=n}(x) \leq \frac{81\cdot 2^{-R_A-R_B + \dseps{\eta}(p_{YN} \| p_Y \times p_N)}}{\delta^5}\bigg\}\nonumber\\
&& \leq \Pr_{(x,n)\leftarrow p_{XN}}\bigg\{p_{X \mid N=n}(x) \leq \frac{9\cdot 2^{-R_A}}{\delta^3} \quad\text{or}\quad p_{X \mid N=n}(x) \leq \frac{81\cdot 2^{-R_A-R_B + \dseps{\eta}(p_{YN} \| p_Y \times p_N)}}{\delta^5}\bigg\} \nonumber\\
&& + \Pr_{(x,n)\leftarrow p_{XN}}\bigg\{\frac{p_{XN}(x,n)}{p_X(x)p_N(n)} \leq \frac{9\cdot 2^{-R_B + \dseps{\eta}(p_{YN} \| p_Y \times p_N)}}{\delta^3}\bigg\}.
\end{eqnarray}
Let $R_B$ be such that $R_B \geq \dseps{\eta}(p_{YN} \| p_Y \times p_N) + 4\log\frac{2}{\delta}$. Then, we have 
$$\Pr_{(x,n)\leftarrow p_{XN}}\bigg\{\frac{p_{XN}(x,n)}{p_X(x)p_N(n)} \leq \frac{9\cdot 2^{-R_B + \dseps{\eta}(p_{YN} \| p_Y \times p_N)}}{\delta^3}\bigg\} \leq \Pr_{(x,n)\leftarrow p_{XN}}\bigg\{\frac{p_{XN}(x,n)}{p_X(x)p_N(n)} \leq \delta\bigg\} \leq \delta.$$
Moreover, 
$$p_{X \mid N=n}(x) \leq \frac{81\cdot 2^{-R_A-R_B + \dseps{\eta}(p_{YN} \| p_Y \times p_N)}}{\delta^5} \leq \frac{6\cdot 2^{-R_A}}{\delta}.$$ Thus, Equation~\eqref{eq:helperinequalities} is upper bounded by
$$\Pr_{(x,n)\leftarrow p_{XN}}\bigg\{p_{X \mid N=n}(x) \leq \frac{6\cdot 2^{-R_A}}{\delta}\bigg\} + \delta \leq \eps.$$
This completes the achievability proof.

\vspace{0.1in}

\noindent {\bf Converse:} Let $M$ be the message sent by Alice and $N$ be the message sent by Bob. Charlie uses $(M,N)$ to output a $\hat{X}$ such that $\Pr\{X\neq \hat{X}\} \leq \eps$. Let $U$ be uniform random variable over Alice's message and $V$ be uniform random variable over Bob's message. Define a random variable $X^*$ jointly correlated with $X$ as 
$$p_{X^*\mid X=x}(x')= 1 \text{ iff } x'=x.$$
Along the lines similar to Equation \ref{errguarantee}, it holds that
\begin{eqnarray}
\label{eq:XXstartell1}
&&\frac{1}{2}\|p_{X\hat{X}N}-p_{XX^*N}\| = \frac{1}{2}\sum_{x,n}p_{XN}(x,n)\|p_{\hat{X}\mid X=x N=n} - p_{X^*\mid X=x}\|\nonumber\\&& =  \frac{1}{2}\sum_{x,n}p_{XN}(x,n)\left(\sum_{x'\neq x} p_{\hat{X}\mid X=x N=n}(x') + 1- p_{\hat{X}\mid X=x N=n}(x)\right) \nonumber \\
&& = \sum_{x,n}p_{XN}(x,n)\left(\sum_{x'\neq x} p_{\hat{X}\mid X=x N=n}(x')\right) \nonumber\\
&& = \Pr\{X\neq \hat{X}\} \leq \eps.
\end{eqnarray}
Consider the inequality
$$p_{XMN}(x,m,n) \leq 2^{R_A}p_{XN}(x,n)p_U(m).$$
By data processing inequality for Bob's operation, it holds that
$$p_{X\hat{X}N}(x,x',n) \leq 2^{R_A}p_{XN}(x,n)p_{W\mid N=n}(x'),$$
for some distribution $(W \mid N=n)$.
From Equation \ref{eq:XXstartell1} and Fact \ref{closeprob}, we conclude that
$$\Pr_{(x,x',n)\drawn p_{XX^*N}}\left\{\frac{p_{XX^*\mid N=n}(x,x')}{p_{X\mid N=n}(x)p_{W\mid N=n}(x')} \geq \frac{2^{R_A}}{\delta}\right\} \leq \frac{\eps}{1-\delta}.$$
Since $x=x'$ for all $(x,x')\drawn p_{XX^*}$, we simplify above to
$$\Pr_{(x, n)\drawn p_{XN}}\left\{\frac{p_{X\mid N=n}(x)}{p_{X\mid N=n}(x)p_{W\mid N=n}(x)} \geq \frac{2^{R_A}}{\delta}\right\} \leq \frac{\eps}{1-\delta}.$$
Combining with the identity
$$\Pr_{(x,n)\drawn p_{XN}}\left\{p_{W\mid N=n}(x) \geq \frac{1}{\delta}p_{X\mid N=n}(x)\right\}\leq \delta,$$
we obtain
\begin{equation}
\label{eq:XgivenNbound}
\Pr_{(x, n)\drawn p_{XN}}\left\{\frac{1}{p_{X\mid N=n}(x)} \geq \frac{2^{R_A}}{\delta^2}\right\} \leq \frac{\eps}{1-\delta} + \delta.
\end{equation}
To bound $R_B$, we proceed as follows. Consider the inequality
$$p_{YN}(y,n) \leq 2^{R_B}p_Y(n)p_V(n).$$ Using the identity 
$$\Pr_{n\drawn p_N}\left\{p_V(n) \geq \frac{1}{\eta} p_N(n)\right\}\leq \eta,$$
we conclude that
$$\Pr_{(y,n)\drawn p_{YN}}\left\{p_{YN}(y,n) \geq \frac{2^{R_B}}{\eta}p_Y(n)p_N(n)\right\}\leq \eta.$$
Thus, 
$$R_B \geq \dseps{\eta}(p_{YN} \| p_Y \times p_N) - \log\frac{1}{\eta}.$$
Along with Equation \ref{eq:XgivenNbound}, this concludes the proof.
\end{proofof}

\section{Optimality of the protocol for Task \ref{task1}}
\label{sec:classicalcompression}

The aim of this section is to relate our achievability result \ref{extensionachievability} with the result of Braverman and Rao~\cite{BravermanRao11}. It may be noted that Braverman and Rao were considering expected communication cost, whereas we are considering the worst case communication cost for Task \ref{task1}. Thus, we have restated the result below accordingly.
  
\begin{theorem}[Braverman and Rao protocol, \cite{BravermanRao11}]
\label{brraotask1}
Let $\eps, \delta \in (0,1)$. Let $R$ be a natural number such that,
\begin{equation*}
R \geq \inf_{p_{N \mid Y}}\dseps{\eps}(p_{XMY}\| p_{Y}(p_{X\mid Y}\times p_{N\mid Y})) + 2\log \frac{1}{\delta}  ,
\end{equation*}
where $(Y,N)\sim p_{Y N}$. There exists a shared randomness assisted protocol in which Alice communicates $R$ bits to Bob and Bob outputs random variable $\hat{M}$ satisfying $\frac{1}{2}\|p_{XYM} - p_{XY\hat{M}}\|\leq \eps+3\delta$. 
\end{theorem}

\begin{proof}
Let $N$ be the random variable that achieves the optimization above. Define 

\begin{eqnarray}
\label{cdefinition}
c &:=& \dseps{\eps}(p_{XMY}\| p_Y(p_{X\mid Y}\times p_{N\mid Y})) \nonumber\\ &=& \min\left\{a : \Pr_{(x,m,y)\drawn p_{XMY}}\bigg\{\frac{p_{XMY}(x,m,y)}{p_Y(y)p_{X\mid Y=y}(x)\cdot p_{N\mid Y = y}(m)}>2^{a}\bigg\}\leq \eps\right\}\nonumber \\ &=& \min\left\{a : \Pr_{(x,m,y)\drawn p_{XMY}}\bigg\{\frac{p_{M\mid X=x}(m)}{p_{N\mid Y=y}(m)}>2^{a}\bigg\}\leq \eps\right\}
\end{eqnarray}
where the last equality follows because $M-X-Y$. Further, define
$$\eps_{x,y}:=  \Pr_{(m)\drawn p_{M\mid X=x}}\bigg\{\frac{p_{M\mid X=x}(m)}{p_{N\mid Y=y}(m)}>2^{c}\bigg\}.$$ It holds that
$\sum_{x,y}p_{XY}(x,y)\eps_{x,y}\leq \eps$.

Let $K$ be the smallest integer such that $K p_{M\mid X=x}(m), K p_{N\mid Y=y}$ are integers. This can be assumed to hold with arbitrarily small error. Further, define the set $\cK:=\left\{1,\cdots, K\right\}$. Let $U$ be a random variable taking values uniformly in $\cM\times \cK$. Define the following \textit{function}: 

\begin{align*}
&f_{NE\mid Y=y} (m,e)  :=
  \begin{cases}
       1~ &\hspace{-4mm}~ \mbox{if } e< K\cdot 2^c\cdot p_{N\mid Y=y}(m)  ,\\
    0 &\hspace{-8mm} \quad \text{ otherwise,}
  \end{cases}
\end{align*}
the following probability distribution:
\begin{align*}
&p_{ME\mid X=x} (m,e)  =
  \begin{cases}
       \frac{1}{K}~ &\hspace{-4mm}~ \mbox{if } e< K\cdot p_{M\mid X=x}(m)  ,\\
    0 &\hspace{-8mm} \quad \text{ otherwise,}
  \end{cases}
\end{align*}
and the following \textit{sub-normalized probability distribution}:
\begin{align*}
&p_{M'E\mid X=x, Y=y} (m,e)  =
  \begin{cases}
       \frac{1}{K}~ &\hspace{-4mm}~ \mbox{if } e< K\cdot \text{min}(p_{M\mid X=x}(m), 2^c\cdot p_{N\mid Y=y}(m))  ,\\
    0 &\hspace{-8mm} \quad \text{ otherwise.}
  \end{cases}
\end{align*}

\noindent {\bf The protocol:} Alice and Bob share $\frac{|M|}{\delta}$ i.i.d. copies of $U$, denoted $\{U_1, U_2, \ldots U_{\frac{|M|}{\delta}}\}$. They also share $\frac{2^c}{\delta^2}$ random hash functions $H_{\ell}: \{1,2,\ldots \frac{|M|}{\delta}\} \rightarrow \{0,1\}$. Upon observing $x\leftarrow X$, Alice takes samples $(m_i, e_i) \leftarrow U_i$ and accepts the first index $i$ that satisfies $p_{ME\mid X=x}(m_i, e_i) >0$. Upon observing $y\leftarrow Y$, Bob takes samples $(m_j, e_j) \leftarrow U_j$ and locates all the indices $j$ that satisfy $f_{NE\mid Y=y}(m_j,e_j) =1$. Let the set of indices located by Bob be $\cJ$. Alice aborts the protocol if she does not find any $i$. Bob aborts the protocol if $|\cJ|\geq \frac{2^c}{\delta}$. Conditioned on not aborting, Alice samples all the hash functions, and sends to Bob the evaluation $\{H_1(i), H_2(i), \ldots H_{\frac{2^c}{\delta^2}}(i)\}$. Bob evaluates all the hash functions for each index $j\in \cJ$. he aborts if there exist $j,j'\in \cJ$ such that $H_{\ell}(j) = H_{\ell}(j')$ for all $\ell$ or there exists no $j$ such that $H_{\ell}(j) = H_{\ell}(i)$ for all $\ell$. Conditioned on not aborting, he has located the correct index $i$, and considers $(m_i, e_i)$. Conditioned on any abort, Bob considers a sample $(m,e)$ the random variable $U$. He outputs $m$ from the sample he considers. Let the overall output of Bob be distributed according to $\hat{M}$.

\vspace{0.1in}

\noindent {\bf Error analysis:} Let $E_1$ be the event that $i\notin \cJ$. Let $E_2$ be the event that Alice does not find any index $i$ or $|\cJ|\geq \frac{2^c}{\delta}$. Let $E_3$ be the event that there exist $j,j'\in \cJ$ such that $H_{\ell}(j) = H_{\ell}(j')$ for all $\ell$. From Equation~\eqref{cdefinition}, $\Pr\{E_1\mid X=x, Y=y\}\leq \eps_{x,y}$. Moreover, as argued in \cite{BravermanRao11} $\Pr\{E_2\}\leq 2\delta$ and $\Pr\{\neg E_3\}\geq 1-\delta$. Thus, $\Pr\{\neg E_2 \cap \neg E_3\} \geq 1-3\delta$.  Conditioned on the events $\neg E_2 \cap \neg E_3$, Bob has obtained a sample $(m,e)$ distributed according to $p_{M'E\mid X=x, Y=y} + \eps_{x,y}\cdot U$, as he outputs a sample according to the sub-normalized distribution $p_{M'E\mid X=x,Y=y}$ conditioned on event $\neq E_1$ and uniform otherwise.  Now, 
\begin{eqnarray*}
\frac{1}{2}\|p_{ME\mid X=x} - p_{M'E\mid X=x, Y=y} - \eps_{x,y}\cdot U\| &\leq& \frac{1}{2}\|p_{ME\mid X=x} - p_{M'E\mid X=x, Y=y}\| + \frac{\eps_{x,y}}{2} \\ &=& \sum_{(m,e): K\cdot p_{M\mid X=x}(m) \geq e \geq K\cdot\text{min}(p_{M\mid X=x}(m), 2^c\cdot p_{N\mid Y=y}(m))}\frac{1}{2K} + \frac{\eps_{x,y}}{2}\\ &\leq& \sum_{m: p_{M\mid X=x} > 2^c\cdot p_{N\mid Y=y}(m)} \frac{p_{M\mid X=x}(m)}{2} + \frac{\eps_{x,y}}{2}\\ &\leq & \eps_{x,y},
\end{eqnarray*}
where the last inequality follows from Equation~\eqref{cdefinition}. Thus, we conclude that 
\begin{eqnarray*}
\frac{1}{2}\|p_{M\mid X=x} - p_{\hat{M}\mid X=x,Y=y}\| &\leq& \frac{1}{2}\|p_{ME\mid X=x} - p_{\hat{M}E\mid X=x,Y=y}\| \\ &\leq& \frac{\Pr\{\neg E_2 \cap \neg E_3\}}{2}\|p_{ME\mid X=x} - p_{M'E\mid X=x, Y=y} - \eps_{x,y}\cdot U\| + \Pr\{E_2 \cup E_3\} \\ &\leq& (1-3\delta)\eps_{x,y} + 3\delta \leq \eps_{x,y} + 3\delta.
\end{eqnarray*}
This implies that $$\frac{1}{2}\|p_{XMY} - p_{X\hat{M}Y}\| \leq \sum_{x,y}p_{XY}(x,y)\eps_{x,y} + 3\delta \leq \eps + 3\delta.$$ This completes the proof.

\end{proof}

We now compare our result (Theorem \ref{extensionachievability}) with Theorem \ref{brraotask1}. To accomplish this, we first define a series of new quantities and relate them to each other. In what follows, we will use $\cP$ to represent a protocol for the Task\ref{task1} discussed in Section \ref{intro}.

\begin{itemize}
\item $\bOpt^{\eps}$: Let $\cP$ be any shared randomness assisted communication protocol in which Alice and Bob work on their respective inputs $(X,Y)$, and Bob outputs a random variable $\hat{M}$ correlated with $XY$. Let $\cP(X,Y) : = (X,Y,\hat{M})$ represent the output of the protocol. We define $\err(\cP):=\frac{1}{2}\|p_{XYM}-p_{XY\hat{M}}\|$ as the error incurred by the protocol and $\cC(\cP)$ as the communication cost of the protocol. Define $$\Opt^{\eps} : =\min_{\cP: \err(\cP)\leq \eps}\cC(\cP).$$
\item $\bOpt_1^{\eps}$: Let $S$ be the shared randomness in a protocol $\cP$. Note that $S$ is independent of $(X,Y)$. Let $V$ be a random variable such that $Y-(X,S)-V$,  $X-(Y,V,S)-\hat{M}$ and $\frac{1}{2}\|p_{XYM}-p_{XY\hat{M}}\| \leq \eps$, where $\hat{M}$ is output by Bob (as discussed above). The random variable $V$ represents the message generated by Alice to Bob in $\cP$. Define
$$\Opt_1^{\eps} : =\min_{(X,Y,U,S,\hat{M},V)}\dinfty(p_{XSV}\|p_{X}\times p_{S}\times p_{U}),$$ where $U$ is the uniformly distributed random variable taking values over same set as $V$. 
\item $\bBr^{\eps}$: The amount of communication needed by the protocol of Braverman and Rao for Task \ref{task1} is denoted by $\Br^{\eps}$ and formally defined below (see also Theorem \ref{brraotask1}). Let $(Y,N)\sim p_{Y N}$. Define $$\Br^{\eps}:=\inf_{p_{N\mid Y}}\dseps{\eps}(p_{XMY}\| p_Y (p_{X\mid Y}\times p_{N\mid Y})) .$$
\item $\bExt^{\delta,\eps}$: This is similar to the quantity obtained in the result of Theorem \ref{extensionachievability} by setting $T$ as uniform random variable $U$. Define $$\Ext^{\delta, \eps}:= \min_{E: Y-X-(M,E)}\left(\dseps{\delta}(p_{XME}\| p_{X} \times p_U) - \dzeroseps{\eps}(p_{YME}\| p_{Y} \times p_U) \right).$$ 
\end{itemize}

The following theorem relates all the quantities defined above to each other. This in turn allows us to prove the optimality of our protocol (see Theorem \ref{extensionconverse}) along with the protocol of Braverman and Rao (Theorem \ref{brraotask1}).

\begin{theorem}
\label{main:theo}
Let $M-X-Y$ and $\eps, \delta \in (0,1)$. Then it holds that 
\begin{enumerate}
\item $\Opt^{\eps} \geq \Opt_1^{\eps}$.
\item $\Opt_1^{\eps} \geq \Br^{\eps/(1-\delta)}-\log(\frac{1}{\delta})$.
\item $\Br^{\eps} + 2\log(\frac{1}{\delta}) \geq \Opt^{\eps+3\delta}$.
\item $\Ext^{\delta,\eps} + 4\log(\frac{5}{\delta}) \geq \Opt^{\eps+4\sqrt{\delta}}$
\item $\Br^{\eps} > \Ext^{0,\eps}$.
\item $\Ext^{0,\eps} \geq \dseps{\eps}(p_{XMEY}\| p_Y(p_{X\mid Y}\times p_{ME\mid Y})) \geq \dseps{\eps}(p_{XMY}\| p_Y(p_{X\mid Y}\times p_{M\mid Y})) \geq \Br^{\eps}$.
\end{enumerate}
\end{theorem}
\begin{proof}
We will prove the inequalities in the order they appear in the Theorem. 
\begin{enumerate}
\item In any one-way communication protocol $\cP$ with a shared randomness $S$, Alice produces a message $V \in \cV$ using $(X,S)$, and communicates this to Bob. Notice that for this choice of $V$ we have  $Y-(X,S)-V$. Using the message $V$, shared randomness $S$ and his input $Y$, Bob outputs $\hat{M}$ such that $\frac{1}{2}\|p_{XY\hat{M}}- p_{XYM}\|\leq \eps$ and $X-(Y,V,S)-\hat{M}$. The total number of bits communicated by Alice to Bob is $\cC(\cP)=\log|\cV|.$ The inequality now follows from the relation $\dinfty(p_{XSV}\| p_{X}\times p_{S}\times p_U)\leq \log|\cV|$  (as $p_{XS} = p_X\times p_S$) and the definition of $\Opt_1^{\eps}.$
\item For the random variables $(X,Y, V,S,U)$ as defined in $\Opt_1^{\eps}$, we prove the following:
\begin{eqnarray}
\label{dinftybr}
\dinfty(p_{XSV} \| p_{X} \times p_S\times p_U) &\overset{a} =& \dinfty(p_{YXSV} \| p_{YX} \times p_S \times p_U)\nonumber\\
 &  =& \dinfty(p_Y(p_{XSV \mid Y}) \| p_Y(p_{X \mid Y}\times p_{S} \times p_U)) \nonumber\\
 & \overset{b} \geq& \min_{S'V'} \dinfty(p_Y(p_{XSV \mid Y}) \| p_Y(p_{X \mid Y}\times p_{S'V'})) \nonumber\\
 & \overset{c}\geq&  \min_{p_{N \mid Y}} \dinfty(p_Y(p_{X \hat{M} \mid Y}) \| p_Y(p_{X \mid Y}\times p_{N\mid Y})) \nonumber\\  & =& 
\min_{p_{N \mid Y}} \dinfty(p_{X \hat{M}Y} \| p_Y(p_{X \mid Y}\times p_{N\mid Y})).
\end{eqnarray}
Above, (a) follows from the fact that $Y-X-(S,V)$; (b) follows by minimizing over all random variables $(S',V')$ and (c) follows from Fact \ref{fact:monoprob} . For a fixed $N$, let $\alpha : = \dinfty(p_{X \hat{M}Y} \| p_Y(p_{X \mid Y}\times p_{N\mid Y}))$. From the relation $\eps \geq \frac{1}{2}\|p_{X\hat{M}Y}- p_{XMY}\|$ and Fact \ref{closeprob}, it holds that
$$\Pr_{(x,y,m)\drawn p_{XYM}}\left\{\frac{p_{XYM}(x,y,m)}{p_{XY\hat{M}}(x,y,m)} \geq \frac{1}{\delta}\right\} \leq \frac{\eps}{1-\delta}.$$
This implies $$\dseps{\frac{\eps}{1-\delta}}(p_{XMY}\| p_Y (p_{X\mid Y}\times p_{N\mid Y})) \leq \dinfty(p_{X \hat{M}Y} \| p_Y(p_{X \mid Y}\times p_{N\mid Y})) + \log\frac{1}{\delta}.$$ Combining with Equation~\eqref{dinftybr}, the item concludes.
\item This is a direct consequence of Theorem \ref{brraotask1}.

\item This is a direct consequence of Theorem \ref{extensionachievability} . 

\item Let $N$ be as obtained from the definition of $\Br^{\eps}$. From Theorem \ref{theo:BRExt} below, it holds that there exists a random variable $E$ such that $(X,Y,M, E)$ satisfies $Y-X-(M, E)$ and 
$$\dseps{\eps}(p_{X M Y} \| p_Y(p_{X\mid Y} \times p_{N\mid Y})) \geq  \dseps{0}(p_{XM E} \| p_{X}\times p_{U}) - \dzeroseps{\eps}(p_{YM E} \| p_{Y} \times p_{U}).$$
This concludes the item.
\item Let $E$ achieve the minimum in definition of $\Ext^{0,\eps}$. Let $\alpha:=  \dseps{0}(p_{XME}\| p_{X} \times p_U)$ and $\beta:= \dzeroseps{\eps}(p_{YME}\| p_{Y} \times p_U)$. Let $\cA \subseteq \cY\times \cM\times \cE$ be the set achieving the optimum for $\beta$. For every $(x,y,m,e)$ such that $(y,m,e)\in \cA$, we have 
$$p_{XMEY}(x,m,e,y) = p_{XY}(x,y)p_{ME\mid X}(m,e) \leq 2^{\alpha}p_{X\mid Y}(x)p_Y(y)p_{U}(m,e) \leq 2^{\alpha-\beta}p_{X\mid Y}(x)p_{YME}(y,m,e).$$ Moreover, $\Pr_{p_{XMEY}}\{\cA\} \geq 1-\eps$. Thus, 
$$\dseps{\eps}(p_{XMEY}\| p_Y(p_{X\mid Y}\times p_{ME\mid Y})) \leq \alpha-\beta.$$ This completes the proof.
\end{enumerate}
\end{proof}  
The following theorem shows that the information theoretic quantity obtained in Theorem \ref{extensionachievability} is upper bounded by the information theoretic quantity obtained in Theorem \ref{brraotask1}.
\begin{theorem}
\label{theo:BRExt}
Let $N$ be the optimal random variable appearing in the definition of $\Br^{\eps}$. There exists a random variable $E$ such that $Y-X-(M, E)$ and
\beq
\dseps{\eps}(p_{X M Y} \| p_Y (p_{X\mid Y} \times p_{N\mid Y})) \geq  \dseps{0}(p_{XM E} \| p_{X}\times p_{U}) - \dzeroseps{\eps}(p_{YM E} \| p_{Y} \times p_{U}) , \nonumber
\enq
where $U$ is uniformly distributed over the set over which the random variable pair $(M,E)$ take values.
\end{theorem}   
\begin{proof} The proof is divided in the following steps.

\noindent{\bf Construction of appropriate extension:} Let $K$ be the smallest integer such that $K p_{M\mid X=x}(m)$ is an integer. This can be assumed to hold with arbitrarily small error. Further, let $E$ be a random variable taking values over
the set $\cK:=\left\{1,\cdots, K\right\}$ and jointly distributed with $(X,M)$ as follows: for every $(m,e, x)\in \cM \times \cK \times \cX$, 
\begin{equation}
\label{extension}
p_{XM E}(x,m,e): =   
\begin{cases}
 \frac{p_{X}(x)}{K} & \mbox{if } e \leq Kp_{M \mid X=x}(m), \\
 0    & \mbox{otherwise}.
\end{cases}
\end{equation}
It can be seen that the property $Y-X-(M, E)$ holds. Let $U$ be a uniform random variable distributed over the set $\cM \times \cK.$ Now we can establish the following: 
\begin{align}
\label{D0equal}
\mathrm{D}^0_{s}( p_{XM E} \| p_{X} \times p_{U}) &\overset{a} = \max_{m,x,e} \log \frac{p_{XM E}(x,m,e)}{p_{X}(x) p_{U}(u)} \nonumber\\
& \overset {b} = \log \frac{|\cM|K}{K} \nonumber \\
& = \log |\cM|,
\end{align}
where (a) follows from the definition of $\mathrm{D}_{s}(p_{XM E} \| p_{X}\times p_{U})$; (b) follows from Equation~\eqref{extension} and the fact that $U$ is uniform over the set $\cM \times \cK.$  

\noindent{\bf Lower bounding smooth hypothesis testing divergence:} For brevity, let $$\dstarft:=\dseps{\eps}(p_{X M Y} \| p_Y (p_{X\mid Y} \times p_{N\mid Y})).$$ Define the following set 
\begin{equation*}
\cA := \left\{(y,m,e) \in \cY\times\cM\times\cK: e \leq K 2^{\dstarft} p_{N\mid Y=y}(m)\right\}.
\end{equation*}
We will prove the following 
\begin{align}
\label{hypo1}
\Pr_{p_{Y}\times p_U}\left\{\cA\right\} &= 2^{-\left(\log |\cM|- \dstarft)\right)};\\
\label{hypo2}
\Pr_{p_{MY E}}\left\{\cA\right\} & \geq 1-\eps.
\end{align}
The theorem now follows from the definition of $\dzeroseps{\eps}(p_{YM E}\| p_{Y}\times p_{U})$ and Equations~\eqref{D0equal},\eqref{hypo1},\eqref{hypo2} as follows: 
\begin{align*}
\dzeroseps{\eps}(p_{YM E}\| p_{Y}\times p_{U}) &\geq \log |\cM|- \dstarft \\
&= \mathrm{D}^0_{s}( p_{XM E} \| p_{X} \times p_{U}) -  \dstarft 
\end{align*}
which leads to $$\dstarft \geq \dseps{0}( p_{XM E} \| p_{X} \times p_{U}) - \dzeroseps{\eps}(p_{YM E}\| p_{Y}\times p_{U}).$$
{\bf Proof of Equation~\eqref{hypo1}}: Notice the following 
\begin{align*}
\Pr_{p_{Y}\times p_U}\left\{\cA\right\} &= \sum_{(y,m,e) \in \cA}p_{Y}(y)p_{U}(m,e) \\
& = \sum_{y \in \cY} p_{Y}(y)  \sum _{(m,e) : (y,m,e) \in \cA } \frac{1}{|\cM|K}\\
& = \sum_{(y,m)\in \cY \times \cM} p_{Y}(y) p_{N|Y=y}(m) \frac{ K 2^{\dstarft}}{|\cM|K}\\
& = \frac{ 2^{\dstarft}}{|\cM|}\\
&= 2^{-\left(\log |\cM|- \dstarft\right)}.
\end{align*}
{\bf Proof of Equation~\eqref{hypo2}}. We have the following: 
\begin{align*}
\Pr_{p_{YM E}} \{A\}& = \sum_{x} p_{X}(x)\sum_{(y,m,e) \in \cA}  p_{Y|X = x} (y) p_{M E|X = x}(m,e) \\
& \overset{a}=\sum_{x} p_{X}(x) \sum_{y}p_{Y|X = x}(y) \sum_m\sum_{\substack{e : e \leq Kp_{M|X = x}(m)\\ (y,m,e)\in \cA}} \frac{1}{K}\\
& \overset{b}\geq \sum_{(x,y)}p_{XY}(x,y)\sum_{m:  p_{M|X = x}(m) \leq 2^{\dstarft}p_{N|Y = y}(m)}p_{M|X = x} (m)\\
& \overset{c} \geq 1 -\eps,
\end{align*}
where $a$ follows from Definition~\eqref{extension}, $b$ follows because for every $x$ 
$$\left\{(y,m,e):p_{M|X = x}(m) \leq 2^{\dstarft}p_{N|Y=y}(m) \mbox{ and } e\leq Kp_{M|X = x}(m) \right\} \subseteq \cA,$$
and $c$ follows from the definition of $\dstarft$. This completes the proof.
\end{proof}

\subsection*{Acknowledgment} 

We thank Marco Tomamichel and Mario Berta for helpful discussions. This work was done when A.A. was affiliated to Centre for Quantum Technologies, National University of Singapore. The work is supported by the Singapore Ministry of Education and the National Research Foundation,
also through the Tier 3 Grant “Random numbers from quantum processes” MOE2012-T3-1-009 and NRF RF
Award NRF-NRFF2013-13.

\bibliographystyle{ieeetr}
\bibliography{References}

\appendix

\section{Deferred proofs} \label{sec:deferred}
\begin{proofof}{Fact~\ref{convexcomb}}
Let $c := \dseps{\eps}(p_{XM} \| p_X\times p_W)$. As shown in \cite[Theorem 1]{ABJT18}, there exists a random variable $(X',M')$ such that 
\beq
\label{ratiosmall}
\forall (x,m): ~ \frac{p_{X'M'}(x,m)}{p_{X'}(x)p_{W}(m)} \leq 2^{c+1}, \quad \forall x:~p_{X'}(x) = p_X(x) 
\enq
and 
\begin{eqnarray}
\label{closestate}
\frac{1}{2}\|p_{X'M'}-p_{XM}\| \leq \eps, 
\end{eqnarray}
Let us construct joint random variables $(J',X',M'_1,\ldots, M'_{2^R})$ from $(X',M')$ in a similar fashion as we constructed joint random variables $(J,X,M_1,\ldots, M_{2^R})$ from $(X,M)$. We note from Equation~\eqref{closestate} that,
\beq
\label{l1inter}
\frac{1}{2}\|p_{X'M'_1\ldots M'_{2^R}}-p_{XM_1\ldots M_{2^R}}\| = \frac{1}{2}\|p_{XM}-p_{X'M'}\| \leq \eps.  \\
\enq
Consider,
\begin{align*}
&\mbox{D}\left(p_{X'M'_1\ldots M'_{2^R}} \| p_{X'}\times p_{W_1}\times\ldots \times p_{W_{2^R}} \right)\\
&\overset{a}= \frac{1}{2^{R}} \sum _{j=1}^{2^R}\left(\mbox{D}\left(p_{X'M'_j} \| p_{X'} \times p_{W_j}\right)-\mbox{D}\left(p_{X'M'_j}\times p_{W_1}\times\ldots \times p_{W_{j-1}}\times p_{W_{j+1}}\times \ldots \times p_{W_{2^R}} \| p_{X'M'_1\ldots M'_{2^R}} \right)\right) \\
& \overset{b}\leq \frac{1}{2^{R}} \sum _{j=1}^{2^R}\left(\mbox{D}\left(p_{X'M'_j} \| p_{X'} \times p_{W_j}\right) - \mbox{D} \left(p_{X'M'_j} \| \frac{1}{2^R}p_{X'M'_j} + \left(1- \frac{1}{2^R}\right)p_{X'} \times p_{W_j} \right)\right)\\
& \overset{c} \leq  \log \left(1+ \frac{2^{c+1}}{2^R} \right)\\
&\overset{d}\leq \frac{\delta^2}{4},
\end{align*}
where (a) follows from Fact~\ref{klprop}; (b) follows from Fact~\ref{fact:monorel} and (c) follows from Equation~\eqref{ratiosmall} and (d) follows since $\log (1 + x) \leq x$ for all real $x$ and from choice of $R$. From Fact~\ref{fact:pinsker} we get,
$$ \frac{1}{2}\| p_{X'M'_1\ldots M'_{2^R}}  - p_{X'}\times p_{W_1}\times\ldots p_{W_{2^R}} \| \leq \delta.$$
Along with Equations~\eqref{l1inter},\eqref{ratiosmall} and the triangle inequality for $\ell_1$ distance gives us
$$ \frac{1}{2}\| p_{XM_1\ldots M_{2^R}}  - p_{X}\times p_{W_1}\times\ldots p_{W_{2^R}} \| \leq \eps+\delta.$$
\end{proofof}

\begin{proofof}{Fact~\ref{genconvexcomb}}
Given random variable $(X,M,N)$, we construct a nearby distribution using the procedure from \cite[Theorem 1]{ABJT18}. For an $x\in \mathcal{X}$, let $\mathrm{Good}_x$ be the set of all pairs $(m,n)$ such that  
$$\frac{p_{XMN}(x,m,n)}{p_X(x)p_U(m)p_V(n)} \leq \frac{\delta^2}{24}\cdot 2^{R_1+R_2} \text{ and } \frac{p_{XM}(x,m)}{p_X(x)p_U(m)} \leq \frac{\delta^2}{24}\cdot 2^{R_1} \text{ and } \frac{p_{XN}(x,n)}{p_X(x)p_V(n)} \leq \frac{\delta^2}{24}\cdot 2^{R_2}.$$
Define 
$$\eps_x:= \Pr_{(m,n) \leftarrow p_{MN\mid X=x}}\left\{\frac{p_{MN\mid X=x}(m,n)}{p_U(m)p_V(n)} \geq \frac{\delta^2}{24}\cdot 2^{R_1+R_2} \text{ or } \frac{p_{M\mid X=x}(m)}{p_U(m)} \geq \frac{\delta^2}{24}\cdot 2^{R_1} \text{ or } \frac{p_{N\mid X=x}(n)}{p_V(n)} \geq \frac{\delta^2}{24}\cdot 2^{R_2}\right\}.$$
Then $\sum_x p_X(x)\eps_x \leq \eps$. Define the random variable $(X', M', N')$ as 
$$p_{(M',N')\mid X'=x}(m,n) := p_{(M,N)\mid X=x}(m,n)\id((m,n)\in \mathrm{Good}_x) + \eps_xp_U(m)p_V(n),$$ $$p_{(X',M',N')}(x,m,n) := p_X(x)p_{(M',N')\mid X=x}(m,n).$$
We have
\begin{eqnarray}
\label{l2inter}
\frac{1}{2}\|p_{XMN}-p_{X'M'N'}\| &=& \sum_x p_X(x)\frac{1}{2}\|p_{M'N'\mid X=x} - p_{MN\mid X=x}\| \nonumber\\ &\leq& \sum_x p_X(x)\frac{1}{2}\left(\eps_x\|p_U(m)p_V(n)\| + \|p_{MN\mid X=x}-p_{MN\mid X=x}\cdot\id(\mathrm{Good}_x)\| \right) \nonumber\\ &=& \sum_x p_X(x)\frac{1}{2}\left(\eps_x + \|p_{MN\mid X=x}\cdot\id(\neg\mathrm{Good}_x)\| \right)\nonumber\\ &=& \sum_x p_X(x)\eps_x \leq \eps.
\end{eqnarray} 
Moreover, we find
$$p_{M'N'\mid X=x}(m,n) \leq \frac{\delta^2}{24}\cdot 2^{R_1+R_2}p_U(m)p_V(n) + \eps_xp_U(m)p_V(n)\leq \frac{\delta^2}{24}\cdot 2^{R_1+R_2+1}p_U(m)p_V(n).$$
Further,
\begin{eqnarray*}
p_{M'\mid X=x}(m) &=& \sum_np_{MN\mid X=x}(m,n)\id((m,n)\in \mathrm{Good}_x) + \eps_x p_U(m) \\
&\leq& p_{M\mid X=x}(m)\id(\exists n: (m,n)\in \mathrm{Good}_x) + \eps_x p_U(m) \\
&\leq& \frac{\delta^2}{24}\cdot 2^{R_1}p_U(m) + \eps_x p_U(m) \leq \frac{\delta^2}{24}\cdot 2^{R_1+1} p_U(m).
\end{eqnarray*}
Similarly, 
\begin{align*}
p_{V \mid X=x}(n) \leq \frac{\delta^2}{24}\cdot 2^{R_2+1}p_V(n)\,.
\end{align*}
Thus, 
\beq
\label{ratiossmall}
 \frac{p_{X'M'}(x,m)}{p_{X'}(x)p_{U}(m)} \leq \frac{\delta^2}{12}\cdot 2^{R_1}, \quad  \frac{p_{X'N'}(x,n)}{p_{X'}(x)p_{V}(n)} \leq \frac{\delta^2}{12}\cdot 2^{R_2}, \quad   \frac{p_{X'M'N'}(x,m,n)}{p_{X'}(x)p_{U}(m)p_{V}(n)} \leq \frac{\delta^2}{12}\cdot 2^{R_1+R_2}, \quad p_{X'}(x)=p_X(x).
\enq
\suppress{Let $\good_x$ be the set of all $(m,n)$ such that 
$$\frac{p_{XM}(x,m)}{p_X(x)p_U(m)} \leq \delta^3\cdot 2^{R_1} \quad, \quad \frac{p_{XN}(x,n)}{p_X(x)p_V(n)} \leq \delta^3\cdot 2^{R_2} \quad ,\quad \frac{p_{XMN}(x,m,n)}{p_X(x)p_U(m)p_V(n)} \leq \delta^3\cdot 2^{R_1+R_2}.$$
Let $\eps_x\defeq 1-\Pr_{p_{MN\mid X=x}}\left\{\good_x\right\}$ and $\good:= \{x: \eps_x \leq 1-\delta\}.$ Similar to Equation~\eqref{largegamma}, we have $$\gamma := \sum_{x\in \good, (m,n)\in \good_x}p_{XMN}(x,m,n) \geq 1-\frac{\eps}{1-\delta}.$$
 Let us define joint random variables $(X',M',N')$ as follows:
\begin{align*}
&p_{X'M'N'} (x,m,n)  =
  \begin{cases}
       \frac{p_{XMN}(x,m,n)}{\gamma}~ &\hspace{-4mm}~ \mbox{if } x\in \good, (m,n) \in \good_x ,\\
    0 &\hspace{-8mm} \quad \text{ otherwise.}
  \end{cases}
\end{align*}
It holds that $p_{X'}(x) =  \frac{p_X(x)\sum_{m\in \good_x}p_{M\mid X=x}(m)}{\gamma}\mathbb{I}(x\in \good)$, where $\mathbb{I}$ is the indicator function. Thus, for all $x\in \good$
$$p_{X'}(x) = \frac{p_X(x)\sum_{m\in \good_x}p_{M\mid X=x}(m)}{\gamma} = \frac{p_X(x)(1-\eps_x)}{\gamma} \geq \frac{\delta}{\gamma}p_X(x).$$ Thus, in a manner similar to Equation~\eqref{ratiosmall}, $\forall (x,m,n):$
\beq
\label{ratiossmall}
 \frac{p_{X'M'}(x,m)}{p_{X'}(x)p_{U}(m)} \leq \delta^2\cdot 2^{R_1} \quad , \quad  \frac{p_{X'N'}(x,n)}{p_{X'}(x)p_{V}(n)} \leq  \delta^2\cdot 2^{R_2} \quad , \quad \frac{p_{X'M'N'}(x,m,n)}{p_{X'}(x)p_{U}(m)p_{V}(n)} \leq \delta^2\cdot 2^{R_1+R_2}.
\enq
and 
\begin{eqnarray}
\label{closestates}
\frac{1}{2}\|p_{X'M'N'}-p_{XMN}\| =1-\gamma \leq \frac{\eps}{1-\delta} \leq \eps+2\eps\delta.
\end{eqnarray}
Let us construct joint random variables $\left(J',K',X',M'_1,\ldots, M'_{2^{R_1}},N'_1,\ldots, N'_{2^{R_2}}\right)$ from $(X',M',N')$ in the same way as we constructed $\left(J,K,X,M_1,\ldots, M_{2^{R_1}},N_1,\ldots, N_{2^{R_2}}\right)$ from $(X,M,N)$. We note that,
\beq
\label{l2inter}
\frac{1}{2}\|p_{X'M'_1\ldots M'_{2^{R_1}}N'_1\ldots N'_{2^{R_2}}} -p_{XM_1\ldots M_{2^{R_1}}N_1\ldots N_{2^{R_2}}}\| = \frac{1}{2}\|p_{XMN}-p_{X'M'N'}\| \leq \eps + 2\eps\delta.
\enq
}
For notational convenience define,
\begin{align*}
\forall j \in [2^{R_1}]: \quad p_{U_{-j}} &:= p_{U_1}\times\ldots \times p_{U_{j-1}}\times p_{U_{j+1}}\times \ldots \times p_{U_{2^{R_1}}} , \\
\forall k \in [2^{R_2}]: \quad  p_{V_{-k}} &:= p_{V_1}\times\ldots \times p_{V_{k-1}}\times p_{V_{k+1}}\times \ldots \times p_{V_{2^{R_2}}} ,\\
q_{X'M'N'} &:= \frac{1}{2^{R_1+R_2}}p_{X'M'_jN'_k}  + \frac{1}{2^{R_1}}\left(1-\frac{1}{2^{R_2}}\right)p_{X'M'_j}\times p_{V_k}  \\ 
& \quad \quad + \frac{1}{2^{R_2}}\left(1-\frac{1}{2^{R_1}}\right)p_{X'N'_k}\times p_{U_j}+\left(1 - \frac{2^{R_1}+2^{R_2}-1}{2^{R_1+R_2}}\right)p_{X'} \times p_{U_j}\times p_{V_k} .
\end{align*}
Consider,
\begin{align*}
&\mbox{D}\left(p_{X'M'_1\ldots M'_{2^{R_1}}N'_1\ldots N'_{2^{R_2}}} \| p_{X'}\times p_{U_1}\times\ldots \times p_{U_{2^{R_1}}}\times p_{V_1}\times\ldots \times  p_{V_{2^{R_2}}} \right)\\
&\overset{a}= \frac{1}{2^{R_1+R_2}} \sum _{j,k}\left(\mbox{D}\left(p_{X'M'_jN'_k} \| p_{X'} \times p_{U_j}\times p_{V_k}\right)-\mbox{D}\left(p_{X'M'_jN'_k}\times  p_{U_{-j}} \times  p_{V_{-k}}  \| p_{X'M'_1\ldots M'_{2^{R_1}} N'_1\ldots N'_{2^{R_2}}} \right)\right)\\
&\overset{b} \leq \frac{1}{2^{R_1+R_2}} \sum _{j,l}\bigg(\mbox{D}\left(p_{X'M'_jN'_k} \| p_{X'} \times p_{U_j}\times p_{V_k}\right)  -  \mbox{D} \bigg(p_{X'M'_jN'_k} \|  q_{X'M'N'} \bigg)\bigg) \\
& \overset{c}\leq  \log \left(1+ \frac{\delta^2\cdot 2^{R_1+R_2}}{12\cdot 2^{R_1+R_2}} + \frac{\delta^2\cdot 2^{R_1}}{12\cdot 2^{R_1}} + \frac{\delta^2\cdot 2^{R_2}}{12\cdot 2^{R_2}} \right)\\
&\overset{d}\leq \frac{\delta^2}{4},
\end{align*}
where (a) follows from Fact~\ref{klprop}; (b) follows from Fact~\ref{fact:monorel}; (c) follows from Equation~\eqref{ratiossmall} and (d) follows since $\log (1 + x) \leq x$ for all real $x$ and from choice of parameters. From Fact~\ref{fact:pinsker} this implies 
$$ \frac{1}{2}\| p_{X'M'_1\ldots M'_{2^{R_1}}N'_1\ldots N'_{2^{R_2}}} - p_{X}\times p_{U_1}\times\ldots \times p_{U_{2^{R_1}}}\times p_{V_1}\times\ldots \times  p_{V_{2^{R_2}}} \| \leq \delta,$$
where we have used $p_{X}=p_{X'}$. This along with Equation~\eqref{l2inter} and the triangle inequality for $\ell_1$ distance gives us the desired.

Now, we show that the choice of $R_1, R_2$ in Equation~\eqref{choicebipconv} suffices. Define,
\begin{align*}
c_1 &:= \dseps{\eps_1}(p_{XM} \| p_X\times p_U), c_2:= \dseps{\eps_2}(p_{XN} \| p_X\times p_V) , c_3 := \dseps{\eps_3}(p_{XMN} \| p_X\times p_U\times p_V)\\
\good_{x,1} &:= \left\{(m,n): \frac{p_{XMN}(x,m,n)}{p_{X}(x)p_{U}(m)p_{V}(n)} \leq 2^{c_1}\right\}, \\ 
\good_{x,2} &:= \left\{(m, n): \frac{p_{XM}(x,m)}{p_{X}(x)p_{U}(m)} \leq 2^{c_2}\right\} ,\\
\good_{x,3} &:= \left\{(m, n): \frac{p_{XN}(x,n)}{p_{X}(x)p_{V}(n)} \leq 2^{c_3}\right\}.
\end{align*}
It can be seen that $\good_{x,1} \cap \good_{x,2} \cap \good_{x,3} \subseteq \good_x$, if the constraints on $R_1, R_2$ are satisfied. This completes the proof.
\end{proofof}

\begin{proofof}{Fact~\ref{fact:position}}
Let $\cA \subseteq \cY \times \cM$ be such that $\Pr_{p_{YM}} \left\{\cA \right\} \geq 1 - \eps$, and
\begin{equation*}
c:= \dzeroseps{\eps}(p_{YM} \| p_Y \times p_W) = - \log \Pr_{p_Y \times p_W} \left\{\cA \right\}.
\end{equation*}
The decoding procedure is to check for each index $j'\in [2^R]$ (in increasing order, for example), whether $(Y, M_{j'}) \in \cA$. Output $J'$ is the first index where this check succeeds.  For the arguments below, let us condition on the event $J=j$ for some fixed $j \in [2^R]$. Let $E_{k}$ be the event that $(Y, M_{j'})\notin \cA$ for all $j'< k$. Abbreviate $Z_0 = (Y,M_1, \ldots M_{2^R})$, and let $Z_k$ be the random variable $Z_0$ conditioned on $E_k$. We have $$\Pr_{p_{Z_0}}\{\neg E_k\} \leq \Pr_{p_{Z_0}}\{(Y, M_{k-1})\in \cA\} + \Pr_{p_{Z_0}}\{\neg E_{k-1}\} = 2^{-c} + \Pr_{p_{Z_0}}\{\neg E_{k-1}\}.$$ Thus, $\Pr_{p_{Z_0}}\{E_k\} \geq 1- k\cdot 2^{-c} \geq 1- 2^{R-c} = 1- \delta$. This implies that $$\frac{1}{2}\|p_{Z_k} - p_{Z_0}\| \leq 1- \Pr_{p_{Z_0}}\{E_k\} \leq \delta.$$ We have $$\frac{1}{2}\|p_{YM_{J'}} - p_{YM}\| \leq \Pr_{p_{Z_0}}\{\neg E_j\} + \Pr_{p_{Z_0}}\{E_j\}\cdot \frac{1}{2}\|p_{YM_{J'} \mid E_j}  - p_{YM}\| \leq \delta + \frac{1}{2}\|p_{YM_{J'} \mid E_j}  - p_{YM}\|.$$ 
To evaluate the second term, we consider the test $(Y, M_j) \in \cA$ performed on $Z_j$. If the test succeeds, then $J'=j$, else $J' >j$. Let $Y'M'$ be the random variable output if the test were performed on $Z_0$. We have $\frac{1}{2}\|p_{YM_{J'} \mid E_j} - p_{Y'M'}\| \leq \frac{1}{2}\|p_{Z_k} - p_{Z_0}\| \leq \delta$. Further, $p_{Y'M'} = \Pr_{p_{YM}}\{\cA\}p_{YM\mid \cA} + \Pr_{p_{YM}}\{\neg A\}p_{Y''M''}$, where $Y''M''$ is an arbitrary random variable if the test fails. Thus $$\frac{1}{2}\|p_{Y'M'} - p_{YM}\| \leq \frac{1}{2}\|\Pr\{\cA\}p_{YM\mid \cA} - p_{YM}\| + \frac{1}{2}\Pr\{\neg A\}\|p_{Y''M''}\| = \Pr_{p_{YM}}\{\neg A\} \leq \eps.$$
Thus, $$\frac{1}{2}\|p_{YM_{J'}} - p_{YM}\| \leq \delta + \frac{1}{2}\|p_{YM_{J'} \mid E_j}  - p_{YM}\| \leq 2\delta + \frac{1}{2}\|p_{Y'M'} - p_{YM}\| \leq \eps + 2\delta.$$
Finally, consider
\begin{align*}
\Pr\{ J' \neq j\} &\leq \Pr\left\{(Y,M_j) \notin \cA\right\} + \Pr\left\{(Y,M_{j'})\in \cA ~ \mbox{for some } j'  \neq j\right\} \\
& \leq \eps + 2^{R} \cdot 2^{- c} \leq \eps +\delta.  
\end{align*}
Therefore,
$$ \Pr\{ J \neq J'\} = \sum_{j \in [2^R]}  \Pr\{J=j\} \cdot \Pr\{ J' \neq j ~|~ J=j\} \leq  \eps + \delta . $$
This completes the proof.
\end{proofof}

\begin{proofof}{Fact~\ref{fact:biposition}}
Let $\cA \subseteq \cY\times \cM \times \cN$ be such that for all $(y,m,n) \in \cA$, 
$$\frac{p_{YMN}(y,m,n)}{p_U(m)p_{YN}(y,n)} \geq \frac{2^{R_1}}{\delta} \quad\text{and}\quad \frac{p_{YMN}(y,m,n)}{p_{YM}(y,m)p_V(n)} \geq \frac{2^{R_2}}{\delta} \quad\text{and}\quad\frac{p_{YMN}(y,m,n)}{p_Y(y)p_U(m)p_V(n)} \geq \frac{2^{R_1+R_2}}{\delta}.$$ 
From the choice of $R_1,R_2$, we have $\Pr_{p_{YMN}}\{\cA\} \geq 1-\eps$. The decoding procedure is to check for each pair $(j', k')\in [2^{R_1}]\times [2^{R_2}]$ (in some lexicographical order), whether $(Y, M_{j'}, N_{k'}) \in \cA$. Output $(J', K')$ is the first pair where this check succeeds.  For the arguments below, let us condition on the event $J=j, K=k$ for some fixed $(j, k)\in [2^{R_1}]\times [2^{R_2}]$. Let $E_{g,h}$ be the event that $(Y, M_{j'}, N_{k'})\notin \cA$ for all $(j', k') < (g,h)$, where the symbol $<$ is as given in the lexicographic ordering. Abbreviate $Z_0 = (Y, M_1, \ldots M_{2^{R_1}}, N_1, \ldots N_{2^{R_2}})$, and let $Z_{g,h}$ be the random variable $Z_0$ conditioned on $E_{g,h}$. We have $$\Pr_{p_{Z_0}}\{\neg E_{g,h}\} \leq 3\delta,$$ by the choice of $R_1, R_2$. This implies that $$\frac{1}{2}\|p_{Z_{g,h}} - p_{Z_0}\| \leq \Pr_{p_{Z_0}}\{E_{g,h}\} \leq 3\delta.$$ We have $$\frac{1}{2}\|p_{YM_{J'}N_{K'}} - p_{YMN}\| \leq \Pr_{p_{Z_0}}\{\neg E_{j,k}\} + \Pr_{p_{Z_0}}\{E_{j,k}\}\cdot \frac{1}{2}\|p_{YM_{J'}N_{K'} \mid E_{j,k}}  - p_{YMN}\| \leq 3\delta + \frac{1}{2}\|p_{YM_{J'}N_{K'} \mid E_{j,k}}  - p_{YMN}\|.$$ 
To evaluate the second term, we consider the test $(Y, M_j, N_k) \in \cA$ performed on $Z_{j,k}$. If the test succeeds, then $(J', K')=(j,k)$, else $(J', K') >(j,k)$. Let $Y'M'N'$ be the random variable output if the test were performed on $Z_0$. We have $\frac{1}{2}\|p_{YM_{J'}N_{K'} \mid E_{j,k}} - p_{Y'M'N'}\| \leq \frac{1}{2}\|p_{Z_{j,k}} - p_{Z_0}\| \leq 3\delta$. Further, $p_{Y'M'N'} = \Pr_{p_{YMN}}\{\cA\}p_{YMN\mid \cA} + \Pr_{p_{YMN}}\{\neg A\}p_{Y''M''N''}$, where $Y''M''N''$ is an arbitrary random variable if the test fails. Thus $$\frac{1}{2}\|p_{Y'M'N'} - p_{YMN}\| \leq \frac{1}{2}\|\Pr_{p_{YMN}}\{\cA\}p_{YMN\mid \cA} - p_{YMN}\| + \frac{1}{2}\Pr_{p_{YMN}}\{\neg A\}\|p_{Y''M''N''}\| = \Pr_{p_{YMN}}\{\neg A\} \leq \eps.$$
Thus, $$\frac{1}{2}\|p_{YM_{J'}N_{K'}} - p_{YMN}\| \leq 3\delta + \frac{1}{2}\|p_{YM_{J'}N_{K'} \mid E_j}  - p_{YMN}\| \leq 6\delta + \frac{1}{2}\|p_{Y'M'N'} - p_{YMN}\| \leq \eps + 6\delta.$$
Furthermore, consider,
\begin{align*}
\Pr\{ (J', K')  \neq (j, k)\} &\leq \Pr\left\{(Y,M_j, N_k) \notin \cA\right\} + \Pr\left\{(Y,M_{\tilde{j}},N_{\tilde{k}})\in \cA ~ \mbox{for some } (\tilde{j}, \tilde{k})  \neq (j, k) \right\} \\
& \leq \eps + 2^{R_1}\cdot 2^{-R_1}\delta + 2^{R_2}\cdot 2^{-R_2}\delta+ 2^{R_1 + R_2} \cdot 2^{-R_1-R_2}\delta \leq \eps + 3\delta.  
\end{align*}
Therefore,
\begin{eqnarray*}
\Pr\{ (J, K) \neq (J', K')\} &=& \sum_{(j, k) \in [2^{R_1}] \times [2^{R_2}]}  \Pr\{(J, K)=(j, k)\} \cdot \Pr\{ (J', K') \neq (j, k) ~|~ (J, K)=(j, k)\} \\ &\leq& \eps  + 3\delta. 
\end{eqnarray*}
Now, we show that the choice of $R_1,R_2$ in Equation~\eqref{choicebipartite} suffices. For $i\in \{1,2,3\}$, let $\cA_i \subseteq \cY\times \cM \times \cN$ be such that $\Pr_{p_{YMN}} \left\{\cA_i \right\} \geq 1 - \eps_i$, and
\begin{eqnarray*}
R_1 - 2\log\delta&=& \dzeroseps{\eps_1}(p_{YMN} \| p_U \times p_{YN}) = - \log \Pr_{p_U \times p_{YN}} \left\{\cA_1 \right\}\\
R_2-2\log\delta &=& \dzeroseps{\eps_2}(p_{YMN} \| p_{YM} \times p_V) = - \log \Pr_{p_{YM} \times p_V} \left\{\cA_2 \right\}\\
R_1+R_2-2\log\delta &=& \dzeroseps{\eps_3}(p_{YMN} \| p_Y\times p_U \times p_V) = - \log \Pr_{p_Y\times p_U \times p_V} \left\{\cA_3 \right\}.
\end{eqnarray*}
Let $\cB_i\subseteq \cA_i$, for $i\in \{1,2,3\}$, be defined as follows.
\begin{eqnarray*}
\forall (y,m,n) \in \cB_1 &:& p_{YMN}(y,m,n)\leq \frac{2^{R_1}}{\delta}p_{U}(m)p_{YN}(y,n)\\
\forall (y,m,n) \in \cB_2 &:& p_{YMN}(y,m,n)\leq \frac{2^{R_2}}{\delta}p_{YM}(y,m)p_V(n)\\
\forall (y,m,n) \in \cB_3 &:& p_{YMN}(y,m,n)\leq \frac{2^{R_1+R_2}}{\delta}p_Y(y)p_{U}(m)p_V(n).
\end{eqnarray*}
Then, $\Pr_{YMN}\{\cB_i\} \leq \delta$ for all $i\in \{1,2,3\}$. This implies that $$\Pr_{YMN}\left\{(\cA_1\setminus \cB_1) \cap (\cA_2\setminus \cB_2) \cap (\cA_3\setminus \cB_3)\right\} \geq 1- \eps_1-\eps_2-\eps_3 - 3\delta \geq 1-\eps.$$ Further, for all $(y,m,n) \in (\cA_1\setminus \cB_1) \cap (\cA_2\setminus \cB_2) \cap (\cA_3\setminus \cB_3)$, we have 
$$\frac{p_{YMN}(y,m,n)}{p_{U}(m)p_{YN}(y,n)}\geq \frac{2^{R_1}}{\delta} \quad \text{and}\quad
\frac{p_{YMN}(y,m,n)}{p_{YM}(y,m)p_V(n)}\geq \frac{2^{R_2}}{\delta} \quad \text{and}\quad \frac{p_{YMN}(y,m,n)}{p_Y(y)p_{U}(m)p_V(n)}\geq \frac{2^{R_1+R_2}}{\delta}.$$
This satisfies all the criteria for the set $\cA$, which completes the proof.
\end{proofof}

\end{document}